\documentclass[11pt]{article}
\usepackage{graphicx}
\input{epsf.sty}
\def\Journal#1#2#3#4{{#1} {\bf #2}, #3 (#4)}

\def\NCA{\em Nuovo Cimento}

\def\NPB{{\em Nucl. Phys.} B}
\def\PLB{{\em Phys. Lett.}  B}
\def\PRL{\em Phys. Rev. Lett.}
\def\PRD{{\em Phys. Rev.} D}
\def\ZPC{{\em Z. Phys.} C}
\def\JHEP{{\em JHEP} }

\def\mco{\multicolumn}

\def\dalemb#1#2{{\vbox{\hrule height .#2pt
        \hbox{\vrule width.#2pt height#1pt \kern#1pt
                \vrule width.#2pt}
        \hrule height.#2pt}}}
\newenvironment{scalepic}[3]
  {\begin{center} \stepcounter{equation}
\def\@currentlabel{\theequation}      
   \label{#3}
   \begin{picture}(350,50)(-40,-25)%
   \put(0,10){\makebox(0,0)[rb]{#1}}
   \put(0,-10){\makebox(0,0)[rt]{#2}}
   \put(0,0){\makebox(-50,0){$\updownarrow$}}
   \put(0,0){\vector(1,0){300}}
   \put(356,0){\makebox(0,0)
} }
  { \end{picture} \end{center} }
\newcommand{\scaleitem}[3]
 { \put(#1,0){ \put(0,-5){\line(0,1){10}}
               \put(0,10){\makebox(0,0)[b]{#2}}
               \put(0,-10){\makebox(0,0)[t]{#3}} }}

\newcommand{\be}{\begin{equation}}
\newcommand{\ee}{\end{equation}}
\newcommand{\beba}{\begin{equation}\begin{array}{lcl}}
\newcommand{\eaee}{\end{array}\end{equation}}
\newcommand{\bea}{\begin{eqnarray}}
\newcommand{\eea}{\end{eqnarray}}
\newcommand{\ba}{\begin{array}}
\newcommand{\ea}{\end{array}}

\def\simlt{\mathrel{\lower2.5pt\vbox{\lineskip=0pt\baselineskip=0pt
           \hbox{$<$}\hbox{$\sim$}}}}
\def\simgt{\mathrel{\lower2.5pt\vbox{\lineskip=0pt\baselineskip=0pt
           \hbox{$>$}\hbox{$\sim$}}}}

\begin{document}
\thispagestyle{empty}
\rightline{\normalsize\sf hep-th/0102202}
\rightline{\normalsize CERN-TH/2001-065}
\rightline{\normalsize 0023/IHP}
\rightline{\normalsize February 2001}
\vskip 0.2truecm

\centerline{\Large\bf String and D-brane Physics at Low Energy$^\dagger$}
\vskip 0.4truecm
\centerline{{\bf I. Antoniadis}$^*$}

\begin{center}
CERN, Theory Division,
           CH-1211 Geneva 23, Switzerland
\end{center}



\noindent
{\bf 1.}~~ Preliminaries\\
{\bf 2.}~~ Heterotic string and motivations for large volume
compactifications\\  
2.1~ Gauge coupling unification\\
2.2~ Supersymmetry breaking by compactification\\
{\bf 3.}~~ M-theory on $S^1/Z_2$``$\times$"Calabi-Yau\\
{\bf 4.}~~ Type I/I$^\prime$ string theory and D-branes\\
4.1~ Low-scale strings and extra-large transverse dimensions\\
4.2~ Relation type I/I$^\prime$ -- heterotic\\
{\bf 5.}~~ Type II theories\\
5.1~ Low-scale IIA strings and tiny coupling\\
5.2~ Large dimensions in type IIB\\
5.3~ Relation type II -- heterotic\\
{\bf 6.}~~ Theoretical implications\\
6.1~ U.V./I.R. correspondence\\
6.2~ Unification \\
6.3~ Supersymmetry breaking and scales hierarchy \\
6.4~ Electroweak symmetry breaking in TeV-scale strings\\
{\bf 7.}~ Scenarios for studies of experimental constraints\\
{\bf 8.}~ Extra-dimensions along the world brane: KK excitations of gauge
bosons\\
8.1~ Production at hadron colliders\\
8.2~ High precision data low-energy bounds\\
8.3~ One extra dimension for other cases\\
8.4~ More than one extra dimension \\
{\bf 9.}~ Extra-dimensions transverse to the brane world: KK excitations
of gravitons\\
9.1~ Signals from missing energy experiments\\
9.2~ Gravity modification and sub-millimeter forces\\
{\bf 10.} Dimension-eight operators and  limits on the string scale\\
{\bf 11.} D-brane Standard Model\\
11.1 Hypercharge embedding and the weak angle\\
11.2 The fate of $U(1)$'s and proton stability\\
{\bf 12.} Appendix: Supersymmetry breaking in type I strings\\
12.1 Scherk-Schwarz deformations\\
12.2 Brane supersymmetry breaking

\vskip 0.1truecm
\noindent
$^\dagger${\small Lectures given at Centre Emile Borel during the semester
``Supergravity, Superstrings and M-theory".}\\
$^*${\small On leave from {\it Centre de Physique
Th{\'e}orique, Ecole Polytechnique, 91128 Palaiseau} (UMR 7644)}

\vfill\eject
\textheight 205mm


\section{Preliminaries}
In critical (ten) dimensions, any consistent superstring theory has two
parameters: a mass (or length) scale $M_s$ ($l_s=M_s^{-1}$), and a
dimensionless string coupling $\lambda_s$ given by the vacuum expectation
value (VEV) of the dilaton field
$e^{<\phi>}=\lambda_s$~\cite{strings,sao}: 
\be
D=10:\qquad\quad M_s=l_s^{-1}\qquad \lambda_s\, .
\label{tenD}
\ee
Upon compactification in $D=4$ dimensions on a compact manifold of volume
$V$, these parameters determine the four-dimensional (4d) Planck mass (or
length) $M_p$ ($l_p=M_p^{-1}$) and the dimensionless gauge coupling $g$
at the string scale. For simplicity, in the following we drop all
numerical factors from our formulae, while, when needed, we use the
numerical values:
\be
D=4:\qquad\quad M_p\simeq 1.2\times 10^{19}\ {\rm GeV}\qquad 
g\simeq 1/5\, .
\label{fourD}
\ee
Moreover, the weakly coupled condition implies that $\lambda_s<<1$. Our
method in the following consists in expressing the 10d parameters
$(M_s,\lambda_s)$ in terms of the 4d ones and the compactification
volume, in heterotic ($s=H$), type I ($s=I$) and type II ($s=II$) string
theories, and then discuss the conditions on possible large volume or low
string scale realizations, keeping the string coupling small.

An important point is that the compactification volume will always be
chosen to be bigger than unity in string units, $V>l_s^6$. This can
be done by a T-duality transformation which exchanges the role of
the Kaluza-Klein (KK) momenta $p$ with the string winding modes $w$. For
instance, in the case of one compact dimension on a circle of radius $R$,
they read:
\be
p={m\over R} \qquad ;\qquad w={nR\over l_s^2}\, ,
\label{pw}
\ee
with integers $m,n$. T-duality inverts the compactification radius and
rescales the string coupling:
\be
R\to{l_s^2\over R}\qquad \lambda_s\to\lambda_s{l_s\over R}\, ,
\label{Tdual}
\ee
so that the lower-dimensional coupling $\lambda_s\sqrt{l_s/R}$ remains
invariant. When $R$ is smaller than the string scale, the winding modes
become very light, while T-duality trades them as KK momenta in terms of
the dual radius ${\tilde R}\equiv l_s^2/R$. The enhancement of the string
coupling is then due to their multiplicity which diverges in the limit
$R\to 0$ (or ${\tilde R}\to\infty$).

\section{Heterotic string and motivations for large volume
compactifications}

In heterotic string, gauge and gravitational interactions appear at the
same (tree) level of perturbation theory (spherical world-sheet
topology), and the corresponding effective action is~\cite{strings,sao}:
\be
S=\int d^4x{V\over\lambda_H^2}(l_H^{-8}{\cal R}+l_H^{-6}F^2)\, ,
\ee
upon compactification in four dimensions. Here, for simplicity, we kept
only the gravitational and gauge kinetic terms, in a self-explanatory
notation. Identifying their respective coefficients with the 4d
parameters $1/l_p^2$ and $1/g^2$, one obtains:
\be
M_H=gM_p\qquad\qquad \lambda_H=g{\sqrt{V}\over l_H^3}\, .
\label{het}
\ee
Using the values (\ref{fourD}), one obtains that the heterotic string
scale is near the Planck mass, $M_H\simeq 10^{18}$, while the string is
weakly coupled when the internal volume is of order of the string scale,
$V\sim l_s^6$. However, despite this fact, there are physical motivations
which suggest that large volume compactifications, and thus strong
coupling, may be relevant in physics~\cite{ia}. These come from gauge
coupling unification and supersymmetry breaking by compactification,
which we discuss below.

\subsection{Gauge coupling unification}

It is a known fact that the three gauge couplings of the Standard Model,
when extrapolated at high energies assuming the particle content of its
$N=1$ minimal supersymmetric extension (MSSM), they meet at an energy
scale $M_{\rm GUT}\simeq 2\times 10^{16}$ GeV. At the one-loop level, one
has:
\be
{1\over g^2_a(\mu)}={1\over g^2}+{b_a\over 4\pi}
\ln{M_{\rm GUT}^2\over\mu^2}\, ,
\ee
where $\mu$ is the energy scale and $a$ denotes the 3 gauge group factors
of the Standard Model $SU(3)\times SU(2)\times U(1)$. The value of
$M_{\rm GUT}$ is very near the heterotic string scale, but it differs by
roughly two orders of magnitude. If one takes seriously this discrepancy,
a possible way to explain it is by introducing large compactification
volume.

Consider for instance one large dimension of size $R$, so that 
$V\sim Rl_H^5$. Identifying $M_{\rm GUT}$ with the compactification scale
$R^{-1}$, this requires $R\sim 100l_H$. Alternatively, one can use string
threshold corrections which grow linearly with $R$~\cite{dkl}. Assuming
that they can account for the discrepancy, one needs roughly 
$R/l_H\sim\ln(M_H^2/M_{\rm GUT}^2)\sim 10$. As a result, the string
coupling (\ref{het}) equals $\lambda_H\sim 0.5 - 2$ which enters in the
strongly coupled regime.

\subsection{Supersymmetry breaking by compactification}

In contrast to ordinary supergravity, where supersymmetry breaking can be
introduced at an arbitrary scale, through for instance the gravitino,
gaugini and other soft masses, in string theory this is not possible
(perturbatively). The only way to break supersymmetry at a scale
hierarchically smaller than the (heterotic) string scale is by
introducing a large compactification radius whose size is set by the
breaking scale. This has to be therefore of the order of a few TeV in
order to protect the gauge hierarchy. An explicit proof exists for
toroidal and fermionic constructions, although the result is believed to
apply to all compactifications~\cite{ablt,kp}. This is one of the very
few general predictions of perturbative (heterotic) string theory that
leads to the spectacular prediction of the possible existence of extra
dimensions accessible to future accelerators~\cite{ia}. The main
theoretical problem is though the strong coupling, as mentioned above.

The strong coupling problem can be understood from the effective field
theory point of view from the fact that at energies higher than the
compactification scale, the KK excitations of gauge bosons and other
Standard Model particles will start being produced and contribute to
various physical amplitudes. Their multiplicity turns very rapidly the
logarithmic evolution of gauge couplings into a power
dependence~\cite{tv}, invalidating the perturbative description, as
expected in a higher dimensional non-renormalizable gauge theory. A
possible way to avoid this problem is to impose conditions which prevent
the power corrections to low-energy couplings~\cite{ia}. For gauge
couplings, this implies the vanishing of the corresponding
$\beta$-functions, which is the case for instance when the KK modes are
organized in multiplets of $N=4$ supersymmetry, containing for every
massive spin-1 excitation, 2 Dirac fermions and 6 scalars. Examples of
such models are provided by orbifolds with no $N=2$ sectors with respect
to the large compact coordinate(s).

The simplest example of a one-dimensional orbifold is an interval of
length $\pi R$, or equivalently $S^1/Z_2$ with $Z_2$ the coordinate
inversion. The Hilbert space is composed of the untwisted sector,
obtained by the $Z_2$-projection of the toroidal states (\ref{pw}), and
of the twisted sector which is localized at the two end-points of the
interval, fixed under the $Z_2$ transformations. This sector is chiral
and can thus naturally contain quarks and leptons, while gauge fields
propagate in the (5d) bulk.

Similar conditions should be imposed to Yukawa's and in principle to
higher (non-renormalizable) effective couplings in order to ensure a soft
ultraviolet (UV) behavior above the compactification scale. We now know
that the problem of strong coupling can be addressed using string
S-dualities which invert the string coupling and relate a strongly
coupled theory with a weakly coupled one~\cite{sao}. For instance, as we
will discuss below, the strongly coupled heterotic theory with one large
dimension is described by a weakly coupled type IIB theory with a tension
at intermediate energies $(Rl_H)^{-1/2}\simeq 10^{11}$ GeV~\cite{ap}.
Furthermore, non-abelian gauge interactions emerge from tensionless
strings~\cite{w95} whose effective theory describes a higher-dimensional
non-trivial infrared fixed point of the renormalization group~\cite{sei}.
This theory incorporates all conditions to low-energy couplings that
guarantee a smooth UV behavior above the compactification scale. In
particular, one recovers that KK modes of gauge bosons form $N=4$
supermultiplets, while matter fields are localized in four dimensions. It
is remarkable that the main features of these models were captured
already in the context of the heterotic string despite its strong
coupling~\cite{ia}.

In the case of two or more large dimensions, the strongly coupled
heterotic string is described by a weakly coupled type IIA or type
I/I$^\prime$ theory~\cite{ap}. Moreover, the tension of the dual string
becomes of the order or even lower than the compactification scale. In
fact, as it will become clear in the following, in the context of any
string theory other than the heterotic, the simple relation (\ref{het})
that fixes the string scale in terms of the Planck mass does not hold and
therefore the string tension becomes an arbitrary parameter~\cite{w}. It
can be anywhere below the Planck scale and as low as a few TeV~\cite{l}.
The main advantage of having the string tension at the TeV, besides its
obvious experimental interest, is that it offers an automatic solution to
the problem of gauge hierarchy, alternative to low-energy supersymmetry or
technicolor~\cite{add,aadd,ab}.

\section{M-theory on $S^1/Z_2$``$\times$"Calabi-Yau}

The strongly coupled $E_8\times E_8$ heterotic string compactified on a
Calabi-Yau manifold (CY) of volume $V$ is described by the 11d M-theory
compactified on an interval $S^1/Z_2$ of length $\pi R_{11}$ times the
same Calabi-Yau~\cite{hw}. Gravity propagates in the 11d bulk, containing
besides the metric and the gravitino a 3-form potential, while gauge
interactions are confined on two 10d boundaries (9-branes) localized at
the two end-points of the interval and containing one $E_8$ factor each.
The corresponding effective action is
\be
S_H=\int d^4x V({R_{11}\over l_M^9}{\cal R}+{1\over l_M^6}F^2)\, .
\label{SH}
\ee
It follows that
\be
l_M=(g^2V)^{1/6}\qquad\qquad R_{11}=g^2{l_M^3\over l_P^2}\, .
\label{Mth}
\ee
The validity of the 11d supergravity regime is when $R_{11}>l_M$ and
$V>l_M^6$ implying $g<1$ by virtue of eq.(\ref{Mth}). Comparison with
the heterotic relations (\ref{het}) yields:
\be
l_M=l_H\lambda_H^{1/3}\qquad\qquad R_{11}=l_H\lambda_H\, ,
\label{M-het}
\ee
which shows in particular that $R_{11}$ is the string coupling in
heterotic units. As a result, at strong coupling $\lambda_H>1$ the M
theory scale and the 11d radius are larger than the heterotic length:
$R_{11}>l_M>l_H$.

Imposing the M-theory scale $l_M^{-1}$ to be at 1 TeV, one finds from the
relations (\ref{Mth}) a value for the radius of the 11th dimension of the
size of the solar system, $R_{11}\simeq 10^8$ kms, which is obviously
excluded experimentally. On the other hand, imposing a value for
$R_{11}\simeq 1$ mm which is the shortest length scale that gravity is
tested experimentally, one finds a lower bound for the M-theory scale
$l_M^{-1}\simgt 10^7$ GeV~\cite{ckm}.

While the relations (\ref{Mth}) seem to impose no theoretical constraint
to $l_M$, there is however another condition to be imposed beyond the
classical approximation~\cite{w}. This is because at the next order the
factorized space $S^1/Z_2\times\rm CY$ is not any more solution of the
11d supergravity equations, which require the size of the Calabi-Yau
manifold to depend on the 11th coordinate $x_{11}$ along the interval.
This can be seen for instance from the supersymmetry transformation of
the 3-form potential (with field-strength $G^{(4)}$) which acquires non
vanishing contributions from the 10d boundaries:
\be
\delta G^{(4)}=l_M^6\delta(x_{11})\left({\rm tr}F\wedge F-
{1\over 2}{\rm tr}{\cal R}\wedge{\cal R}\right)+
\left( x_{11}\leftrightarrow \pi R_{11}-x_{11}, 
F\leftrightarrow F'\right)\, .
\ee
As a result, the volume of CY varies linearly along the interval, to
leading order:
\be
V(x_{11})=V(0)-x_{11}l_M^3 \int_{\rm CY} \omega\wedge\left(
{\rm tr}F'\wedge F'- {\rm tr}F\wedge F\right) \, ,
\label{V}
\ee
where $\omega\sim V^{1/3}$ is the K\"ahler form on the six-manifold CY.

It follows that there is an upper bound on $R_{11}$, otherwise the gauge
coupling in one of the two walls blows up when the volume of CY shrinks
to zero size. Choosing $V(0)\equiv V$ and imposing $V(\pi R)\ge 0$, 
eq.(\ref{V}) yields $R_{11}\simlt V^{2/3}/l_M^3$ and through the
relations (\ref{Mth}):
\be
l_P\simgt g^{5/3}l_M=g^2 V^{1/6}\, .
\ee
This implies a lower bound for the M-theory scale 
$l_M^{-1}\simgt g^{5/3}M_P$, or equivalently for the unification scale
$M_{\rm GUT}\equiv V^{-1/6}\simgt g^2M_P$. Taking into account the
numerical factors, on finds for the lower bound the right order of
magnitude $M_{\rm GUT}\sim 10^{16}$ GeV, providing a solution to the
perturbative discrepancy between the unification and heterotic string
scales, discussed in section 2.1~\cite{w}. Note that this bound
does not hold in the case of symmetric embedding, where one has 
${\rm tr}F'\wedge F'-{\rm tr}F\wedge F=0$ and thus the correction in
eq.(\ref{V}) vanishes.

\section{Type I/I$^\prime$ string theory and D-branes}

In ten dimensions, the strongly coupled $SO(32)$ heterotic string is 
described by the type I string, or upon T-dualities to type
I$^\prime$~\cite{pw,sao}.\footnote{In lower dimensions, type I$^\prime$
theories can also describe a class of M-theory compactifications.} Type
I/I$^\prime$ is a theory of closed and open unoriented strings. Closed
strings describe gravity, while gauge interactions are described by open
strings whose ends are confined to propagate on D-branes. It follows that
the 6 internal compact dimensions are separated into longitudinal
(parallel) and transverse to the D-branes. Assuming that the Standard
Model is localized on a $p$-brane with $p\ge 3$, there are $p-3$
longitudinal and $9-p$ transverse compact dimensions. In contrast to the
heterotic string, gauge and  gravitational interactions appear at
different order in perturbation theory and the corresponding effective
action reads~\cite{strings,sao}:
\be
S_{I}=\int d^{10}x \frac{1}{\lambda_I^2 l_I^8} {\cal R} + 
\int d^{p+1}x \frac{1}{\lambda_I l_I^{p-3}} F^2\, ,
\label{SI}
\ee
where the $1/\lambda_I$ factor in the gauge kinetic terms corresponds to
the disk diagram. 

Upon compactification in four dimensions, the Planck length and gauge
couplings are given to leading order by
\begin{equation}
\frac{1}{l_P^2}=\frac{V_\parallel V_\perp}{\lambda_I^2 l_I^8}\ ,\qquad
\frac{1}{g^2}=\frac{V_\parallel}{\lambda_I l_I^{p-3}}\, ,
\label{I}
\end{equation}
where $V_\parallel$ ($V_\perp$) denotes the compactification volume 
longitudinal (transverse) to the $p$-brane. From the second relation
above, it follows that the requirement of weak coupling $\lambda_I<1$
implies that the size of the longitudinal space must be of order of the
string length ($V_\parallel\sim l_I^{p-3}$), while the transverse volume
$V_\perp$ remains unrestricted. One thus has
\begin{equation}
M_P^2=\frac{1}{g^4 v_\parallel}M_I^{2+n}R_\perp^n\ ,\qquad
\lambda_I =g^2 v_\parallel\, ,
\label{treei}
\end{equation}
to be compared with the heterotic relations (\ref{het}). Here, 
$v_\parallel\simgt 1$ is the longitudinal volume in string units, 
and we assumed an isotropic transverse space of $n=9-p$ compact 
dimensions of radius $R_\perp$.

\subsection{Low-scale strings and extra-large transverse dimensions}

From the relations (\ref{treei}), it follows that the type I/I$^\prime$
string scale can be made hierarchically smaller than the Planck mass at
the expense of introducing extra large transverse dimensions that
interact only gravitationally, while keeping the string coupling
weak~\cite{aadd,st}. The weakness of 4d gravity $M_I/M_P$ is then
attributed to the largeness of the transverse space $R_\perp/l_I$. An
important property of these models is that gravity becomes strong at the
string scale, although the string coupling remains weak. In fact, the
first relation of eq.(\ref{treei}) can be understood as a consequence of
the
$(4+n)$-dimensional Gauss law for gravity, with
\be
G_N^{(4+n)}=g^4 l_I^{2+n}v_\parallel
\label{GN}
\ee
the Newton's constant in $4+n$ dimensions.

To be more explicit, taking the type I string scale $M_I$ to be at 1 TeV,
one finds a size for the transverse dimensions $R_\perp$ varying from
$10^8$ km, .1 mm (10$^{-3}$ eV), down  to .1 fermi (10 MeV) for $n=1,2$,
or 6 large dimensions, respectively. The case $n=1$ corresponds to
M-theory and is obviously experimentally excluded. On the other hand, all
other possibilities are consistent with observations, although barely in
the case $n=2$~\cite{add2}. In particular, sub-millimeter transverse
directions are compatible with  the present constraints from
short-distance gravity measurements which tested Newton's law up to the
cm~\cite{price}. The strongest bounds come from astrophysics and
cosmology and concern mainly the case $n=2$~\cite{add2,COMPTEL}. In fact,
graviton emission during supernovae cooling restricts the 6d Planck scale to
be larger than about 50 TeV, implying $M_I\simgt 7$ TeV, while the graviton
decay contribution to the cosmic diffuse gamma radiation gives even stronger
bounds of about 110 TeV and 15 TeV for the two scales, respectively.

If our brane world is supersymmetric, which protects the hierarchy in the
usual way, the string scale is an arbitrary parameter and can be at
higher energies, in principle up to the Planck scale. However, in the
context of type I/I$^\prime$ theory, the string scale should not be
higher than intermediate energies $M_I\simlt 10^{11}$ GeV, due to the
generic existence of other branes with non supersymmetric world
volumes~\cite{int}. Indeed, in this case, our world would feel the effects
of supersymmetry breaking through gravitationally suppressed interactions
of order $M_I^2/M_P$, that should be less than a TeV. In this context, the
value $M_I\sim 10^{11}$ GeV could be favored, since it would coincide with
the scale of supersymmetry breaking in a hidden sector, without need of
non-perturbative effects such as gaugino condensation. Moreover, the gauge
hierarchy would be minimized, since one needs to introduce transverse
dimensions with size just two orders of magnitude larger than
$l_I$ (in the case of $n=6$) to account for the ratio
$M_I/M_P\simeq 10^{-8}$, according to eq.(\ref{treei}). Note also that
the weak scale $M_W\sim M_I^2/M_P$ becomes T-dual to the Planck scale.

\subsection{Relation type I/I$^\prime$ -- heterotic}

We will now show that the above type I/I$^\prime$ models describe
particular strongly coupled heterotic vacua with large
dimensions~\cite{aq,ap}. More precisely, we will consider the heterotic
string compactified on a 6d manifold with $k$ large dimensions of radius 
$R\gg l_H$ and $6-k$ string-size dimensions and show that for $k\ge 4$ it
has a perturbative type I$^\prime$ description~\cite{ap}. 

In ten dimensions, heterotic and type I theories are related by an
S-duality:
\be
\lambda_I={1\over\lambda_H}\qquad\qquad l_I=\lambda_H^{1/2}l_H\, ,
\label{het-I}
\ee
which can be obtained for instance by comparing eqs.(\ref{het}) with
eqs.(\ref{I}) in the case of 9-branes ($p=9$, $V_\perp=1$,
$V_\parallel=V$). Using from eq.(\ref{het}) that 
$\lambda_H\sim (R/l_H)^{k/2}$, one finds
\be
\lambda_I\sim\left({R\over l_H}\right)^{-k/2}\qquad\qquad
l_I\sim\left({R\over l_H}\right)^{k/4}l_H\, .
\ee
It follows that the type I scale $M_I$ appears as a non-perturbative
threshold in the heterotic string at energies much lower than
$M_H$~\cite{ckm}. For $k<4$, it appears at intermediate energies
$R^{-1}<M_I<M_H$, for $k=4$, it becomes of the order of the
compactification scale $M_I\sim R^{-1}$, while for $k>4$, it appears at
low energies $M_I<R^{-1}$~\cite{aq}. Moreover, since $\lambda_I\ll 1$,
one would naively think that weakly coupled type I theory could describe
the heterotic string with any number $k\ge 1$ of large dimensions.
However, this is not true because there are always some dimensions
smaller than the type I size ($6-k$ for $k<4$ and 6 for $k>4$) and one
has to perform T-dualities (\ref{Tdual}) in order to account for the
multiplicity of light winding modes in the closed string sector, as we
discussed in section 1.1. Note that open strings have no winding
modes along longitudinal dimensions and no KK momenta along transverse
directions. The T-dualities have two effects: (i) they transform the
corresponding longitudinal directions to transverse ones by exchanging KK
momenta with winding modes, and (ii) they increase the string coupling
according to eq.(\ref{Tdual}) and therefore it is not clear that type
I$^\prime$ theory remains weakly coupled.

Indeed for $k<4$, after performing $6-k$ T-dualities on the heterotic
size dimensions, with respect to the type I scale, one obtains a type
I$^\prime$ theory with D($3+k$)-branes but strong coupling:
\be
l_H\to{\tilde l}_H\!=\!{l_I^2\over l_H}\!\sim\!
\left({R\over l_H}\right)^{k/2}l_H
\qquad \lambda_I\to{\tilde\lambda}_I\!=\!
\lambda_I\left({l_I\over l_H}\right)^{6-k}\!\sim\!
\left({R\over l_H}\right)^{k(4-k)/4}\!\!\gg\!\! 1\, .
\label{kl4}
\ee
For $k\ge 4$, we must perform T-dualities in all six internal
directions.\footnote{The case $k=4$ can be treated in the same way, since
there are 4 dimensions that have type I string size and remain inert
under T-duality.} As a result, the type I$^\prime$ theory has D3-branes
with $6-k$ transverse dimensions of radius ${\tilde l}_H$ given in
eq.(\ref{kl4}) and $k$ transverse dimensions of radius 
${\tilde R}=l_I^2/R\sim (R/l_H)^{k/2-1}$, while its coupling remains weak
(of order unity):
\be
\lambda_I\to{\tilde\lambda}_I=\lambda_I
\left({l_I\over l_H}\right)^{6-k}\left({l_I\over R}\right)^k\sim 1\, .
\ee

It follows that the type I$^\prime$ theory with $n$ extra-large
transverse dimensions offers a weakly coupled dual description for the
heterotic string with $k=4,5,6$ large dimensions~\cite{ap}. $k=4$ is
described by $n=2$, $k=6$ (for $SO(32)$ gauge group) is described by
$n=6$, while for $n=5$ one finds a type I$^\prime$ model with 5 large
transverse dimensions and one extra-large. The case $k=4$ is particularly
interesting: the heterotic string with 4 large dimensions, say at a TeV,
is described by a perturbative type I$^\prime$ theory with the string
scale at the TeV and 2 transverse dimensions of millimeter size that are
T-dual to the 2 heterotic string size coordinates. This is depicted in
the following diagram, together with the case $k=6$, where we use
heterotic length units $l_H=1$:
\begin{scalepic}
{H: $k=4$}{I$^\prime$: $n=2$}{scal04}
\scaleitem{30}{$l_H$, $R_{5,6}$}{1}
\scaleitem{140}{$R_{1,2,3,4}=R$}{$l_I$}
\scaleitem{250}{$R^2$}{$\tilde R_{5,6}$}
\end{scalepic}
\begin{scalepic}
{H: $k=6$}{I$^\prime$: $n=6$}{scal06}
\scaleitem{30}{$l_H$}{1}
\scaleitem{140}{$R_{1,\cdots,6}=R$}{}
\scaleitem{195}{$R^{3/2}$}{$l_I$}
\scaleitem{250}{$R^2$}{$\tilde R_{1,\cdots,6}$}
\end{scalepic}

\section{Type II theories}

Upon compactification to 6 dimensions or lower, the heterotic string
admits another dual description in terms of type II (IIA or IIB) string
theory~\cite{ht,sao}. Since in 10 dimensions type II theories have $N=2$
supersymmetry,\footnote{Type IIA (IIB) has two 10d supercharges of
opposite (same) chirality.} in contrast to the heterotic string which has
$N=1$, the compactification manifolds on the two sides should be
different, so that the resulting theories in lower dimensions have the
same number of supersymmetries. The first example arises in 6 dimensions,
where the $E_8\times E_8$ heterotic string compactified on the four-torus
$T^4$ is S-dual to type IIA compactified on the $K3$ manifold that has
$SU(2)$ holonomy and breaks half of the supersymmetries. In lower
dimensions, type IIA and type IIB are related by T-duality (or mirror
symmetry). 

Here, for simplicity, we shall restrict ourselves to 4d compactifications
of type II on $K3\times T^2$, yielding $N=4$ supersymmetry, or more
generally on Calabi-Yau manifolds that are $K3$ fibrations, yielding
$N=2$ supersymmetry. They are obtained by replacing $T^2$ by a ``base"
two-sphere over which $K3$ varies, and they are dual to corresponding
heterotic compactifications on $K3\times T^2$. More interesting
phenomenological models with $N=1$ supersymmetry can be obtained by a
freely acting orbifold on the two sides, although the most general $N=1$
compactification would require F-theory on Calabi-Yau fourfolds, which
is poorly understood at present~\cite{pm}. 

In contrast to  heterotic and type I strings, non-abelian gauge symmetries
in type II models arise non-perturbatively (even though at arbitrarily
weak coupling) in singular compactifications, where the massless gauge
bosons are provided by D2-branes in type IIA (D3-branes in IIB) wrapped
around non-trivial vanishing 2-cycles (3-cycles). The resulting gauge
interactions are localized on $K3$ (similar to a Neveu-Schwarz
five-brane), while matter multiplets would arise from further
singularities, localized completely on the 6d internal space~\cite{kv}.

\subsection{Low-scale IIA strings and tiny coupling}

In type IIA non-abelian gauge symmetries arise in six dimensions from
D2-branes wrapped around non-trivial vanishing 2-cycles of a singular
$K3$.\footnote{Note though that the abelian Cartan subgroup is already in
the perturbative spectrum of the Ramond-Ramond sector.} It follows that
gauge kinetic terms are independent of the string coupling
$\lambda_{IIA}$ and the corresponding effective action is~\cite{sao}:
\be
S_{IIA}=\int d^{10}x \frac{1}{\lambda_{IIA}^2 l_{IIA}^8} {\cal R} + 
\int d^6 x {1\over l_{IIA}^2} F^2\, ,
\label{SIIA}
\ee
which should be compared with (\ref{SH}) of heterotic and (\ref{SI}) of
type I/I$^\prime$. As a result, upon compactification in four dimensions,
for instance on a two-torus $T^2$, the gauge couplings are determined by
its size $V_{T^2}$, while the Planck mass is controlled by the 6d string
coupling $\lambda_{6IIA}$:
\be
\frac{1}{g^2}={V_{T^2}\over l_{IIA}^2} \qquad\qquad
\frac{1}{l_P^2}=\frac{V_{T_2}}{\lambda_{6IIA}^2 l_{IIA}^4}
={1\over\lambda_{6IIA}^2}{1\over g^2l_{IIA}^2}\, .
\label{IIA}
\ee

The area of $T^2$ should therefore be of order $l_{IIA}^2$, while the
string scale is expressed by
\be
M_{IIA}=g\lambda_{6IIA}M_P=
g\lambda_{IIA}M_P{l_{IIA}^2\over\sqrt{V_{K3}}}\, ,
\label{IIA2}
\ee
with $V_{K3}$ the volume of $K3$. Thus, in contrast to the type I
relation (\ref{treei}) where only the volume of the internal six-manifold
appears, we now have the freedom to use both the string coupling and
the $K3$ volume to separate the Planck mass from a string scale, say,
at 1 TeV~\cite{l,ap}. In particular, we can choose a string-size internal
manifold, and have an ultra-weak coupling $\lambda_{IIA}=10^{-14}$ to
account for the hierarchy between the electroweak and the Planck
scales~\cite{ap}. As a result, despite the fact that the string scale is
so low, gravity remains weak up to the Planck scale and string
interactions are suppressed by the tiny string coupling, or equivalently
by the 4d Planck mass. Thus, there are no observable effects in particle
accelerators, other than the production of KK excitations along the two
TeV dimensions of $T^2$ with gauge interactions. Furthermore, the
excitations of gauge multiplets have $N=4$ supersymmetry, even when
$K3\times T^2$ is replaced by a Calabi-Yau threefold which is a $K3$
fibration, while matter multiplets are localized on the base (replacing
the $T^2$) and have no KK excitations, as the twisted states of heterotic
orbifolds.

Above, we discussed the simplest case of type II compactifications with
string scale at the TeV and all internal radii having the string size. In
principle, one can allow some of the $K3$ (transverse) directions to be
large, keeping the string scale low. From eq.(\ref{IIA2}), it follows
that the string coupling $\lambda_{IIA}$ increases making gravity strong
at distances $l_P\sqrt{V_{K3}}/l_{IIA}^2$ larger than the Planck length.
In particular, it becomes strong at the string scale (TeV), when
$\lambda_{IIA}$ is of order unity. This corresponds to 
$V_{K3}/l_{IIA}^4\sim 10^{28}$, implying a fermi size for the four $K3$
compact dimensions.

\subsection{Large dimensions in type IIB}

Above we assumed that both directions of $T^2$ have the string size, so
that its volume is of order $l_{IIA}^2$, as implied by eq.(\ref{IIA}).
However, one could choose one direction much bigger than the string scale
and the other much smaller. For instance, in the case of a rectangular
torus of radii $r$ and $R$, $V_{T^2}=rR\sim l_{IIA}^2$ with
$r\gg l_{IIA}\gg R$. This can be treated by performing a T-duality
(\ref{Tdual}) along $R$ to type IIB: $R\to l_{IIA}^2/R$ and
$\lambda_{IIA}\to\lambda_{IIB}=\lambda_{IIA}l_{IIA}/R$ with
$l_{IIA}=l_{IIB}$. One thus obtains:
\be
\frac{1}{g^2}={r\over R} \qquad\qquad
\frac{1}{l_P^2}=\frac{V_{T_2}}{\lambda_{6IIB}^2 l_{IIB}^4}
={R^2\over\lambda_{6IIB}^2}{1\over g^2l_{IIB}^4}\, .
\label{IIB}
\ee
which shows that the gauge couplings are now determined by the ratio of
the two radii, or in general by the shape of $T^2$, while the Planck mass
is controlled by its size, as well as by the 6d type IIB string coupling.
The string scale can thus be expressed as~\cite{ap}:
\be
M_{IIB}^2=g\lambda_{6IIB}{M_P\over R}\, .
\label{IIB2}
\ee

Comparing these relations with eqs.(\ref{IIA}) and (\ref{IIA2}), it is
clear that the situation in type IIB is the same as in type IIA, unless
the size of $T^2$ is much larger than the string length, $R\gg l_{IIB}$.
Since $T^2$ is felt by gauge interactions, its size cannot be larger than
${\cal O}({\rm TeV}^{-1})$ implying that the type IIB string scale should
be much larger than TeV. From eq.(\ref{IIB2}) and $\lambda_{6IIB}<1$, one
finds $M_{IIB}\simlt\sqrt{M_P/R}$, so that the largest value for the
string tension, when $R\sim 1{\rm TeV}^{-1}$, is an intermediate scale
$\sim 10^{11}$ GeV when the string coupling is of order unity. 

As we will show below, this is precisely the case that describes the
heterotic string with one TeV dimension, which we discussed is section 2.
It is the only example of longitudinal dimensions larger than the string
length in a weakly coupled theory. In the energy range between the KK
scale $1/R$ and the type IIB string scale, one has an effective 6d theory
without gravity at a non-trivial superconformal fixed point described by
a tensionless string~\cite{w95,sei}. This is because in type IIB gauge
symmetries still arise non-perturbatively from vanishing 2-cycles of
$K3$, but take the form of tensionless strings in 6 dimensions, given by
D3-branes wrapped on the vanishing cycles. Only after further
compactification does this theory reduce to a standard gauge theory,
whose coupling involves the shape rather than the volume of the
two-torus, as described above. Since the type IIB coupling is of order
unity, gravity becomes strong at the type IIB string scale and the main
experimental signals at TeV energies are similar to those of type IIA
models with tiny string coupling.

Similar constructions can be also realized in the context of the heterotic
string when the standard model is embedded in non-perturbative gauge group
arising from small instantons. In this case, the heterotic string scale can
also be lowered in the TeV region \cite{yaron}.

\subsection{Relation type II -- heterotic}

We will now show that the above low-scale type II models describe some
strongly coupled heterotic vacua and, in particular, the cases with
$k=1,2,3$ large dimensions that have not a perturbative description in
terms of type I$^\prime$ theory~\cite{ap}. As we described in the
beginning of section 5, in 6 dimensions the heterotic $E_8\times E_8$
superstring compactified on $T^4$ is S-dual to type IIA compactified on
$K3$:
\be
\lambda_{6IIA}={1\over\lambda_{6H}}\qquad\qquad
l_{IIA}=\lambda_{6H}l_H\, ,
\label{het-II}
\ee
which can be obtained, for instance, by comparing eqs.(\ref{IIA}) with
(\ref{het}), using $\lambda_{6H}=\lambda_H l_H^2/\sqrt{V_{T^4}}$.
However, in contrast to the case of heterotic -- type I/I$^\prime$
duality, the compactification manifolds on the two sides are not the same
and a more detailed analysis is needed to study the precise mapping of
$T^4$ to $K3$, besides the general relations (\ref{het-II}).

This can be done easily in the context of M-theory compactified on the
product space of a line interval of length $\pi R_I$ with four circles of
radii $R_1,\cdots$, $R_4$~\cite{op,ap}:
$S^1/Z_2(R_I)\times S^1(R_1)\times T^3(R_2,R_3,R_4)$. One can then
interpret this compactification in various ways by choosing appropriately
one of the radii as that of the eleventh dimension. Considering for
instance $R_I=R_{11}$, one finds the (strongly coupled) heterotic string
compactified on $T^4(R_1,\cdots,R_4)$, while choosing $R_1=R_{11}$, one
finds type IIA compactified on $K3$ of ``squashed" shape
$S^1/Z_2(R_I)\times T^3({\tilde R}_2,{\tilde R}_3,{\tilde R}_4)$, where
the 3 radii ${\tilde R}_i$ will be determined below. In each of the two
cases, one can use the duality relations (\ref{M-het}) to obtain
\be
R_I=\lambda_H l_H=\lambda_{6H}{V_{T^4}^{1/2}\over l_H}
\qquad\qquad R_1=\lambda_{IIA}l_{IIA}
=\lambda_{6IIA}{V_{K3}^{1/2}\over l_{IIA}}\, ,
\ee
while using eqs.(\ref{het-II}) one finds a mapping between the volume of
the internal 4-manifold of one theory and a preferred radius of the
other, in corresponding string units:
\be
{R_I\over l_{IIA}}={V_{T^4}^{1/2}\over l_H^2}\qquad\qquad
{R_1\over l_H}={V_{K3}^{1/2}\over l_{IIA}}\, .
\label{RV}
\ee
The correspondence among the remaining 3 radii can be found, for
instance, by noticing that the S-duality transformations leave invariant
the shape of $T^3$:
\be
{R_i\over R_j}={{\tilde R}_i\over{\tilde R}_j}\qquad\qquad i,j=2,3,4\, ,
\ee
which yields ${\tilde R}_i=l_M^3/(R_jR_k)$ with $i\ne j\ne k\ne i$ and
$l_M^3=\lambda_H l_H^3$. This relation, together with eq.(\ref{RV}),
gives the precise mapping between $T^4$ and $K3$, which completes the
S-duality transformations (\ref{het-II}). We recall that on the type II
side, the four $K3$ directions corresponding to $R_I$ and ${\tilde R}_i$
are transverse to the 5-brane where gauge interactions are localized.

Using the above results, one can now study the possible perturbative type
II descriptions of 4d heterotic compactifications on
$T^4(R_1,\cdots,R_4)\times T^2(R_5,R_6)$ with a certain number $k$ of
large dimensions of common size $R$ and string coupling 
$\lambda_H\sim (R/l_H)^{k/2}\gg 1$. From eq.(\ref{het-II}), the type II
string tension appears as a non-perturbative threshold at energies of the
order of the $T^2$ compactification scale, $l_{II}\sim\sqrt{R_5R_6}$.
Following the steps we used in the context of heterotic -- type I duality,
after T-dualizing the radii which are smaller than the string size, one
can easily show that the $T^2$ directions must be among the $k$ large
dimensions in order to obtain a perturbative type II description. 

It follows that for $k=1$ with, say, $R_6\sim R\gg l_H$, the
type II threshold appears at an intermediate scale
$l_{II}\sim\sqrt{Rl_H}$, together with all 4 directions of $K3$, while
the second, heterotic size, direction of $T^2$ is T-dual (with respect to
$l_{II}$) to $R$: ${\tilde R}_5\equiv l_{II}^2/l_H\sim R$. Thus, one
finds a type IIB description with two large longitudinal dimensions along
the $T^2$ and string coupling of order unity, which is the example
discussed in sections 2.2 and 5.2.
\begin{scalepic}
{H: $k=1$}{IIB, $\lambda\!\!\sim\!\!1$}{scal01}
\scaleitem{40}{$l_H$, $R_{1,\cdots,4}, R_5$}{1}
\scaleitem{110}{$\sqrt{R}$}{$l_{IIB}$, $K3$}
\scaleitem{180}{$R_6=R$}{$T^2({\tilde R}_5,R_6)$}
\end{scalepic}
For $k\ge 2$, the type II scale
becomes of the order of the compactification scale, $l_{II}\sim R$. For
$k=2$, all directions of $K3\times T^2$ have the type II size, while the
type II  string coupling is infinitesimally small, $\lambda_{II}\sim
l_H/R$, which is the example discussed in section 5.1. 
\begin{scalepic}
{H: $k=2$}{II,$\lambda\!\!\sim\!\!1\!/\!R$}{scal02}
\scaleitem{40}{$l_H$, $R_{1,\cdots,4}$}{1}
\scaleitem{180}{$R_{5,6}=R$}{$l_{II}$, $K3$, $T^2(R_{5,6})$}
\end{scalepic}
For $k=3$, 
$l_{II}\sim R_{5,6}\sim R$, while the four (transverse) directions of
$K3$ are extra large: $R_I\sim{\tilde R}_i\sim R^{3/2}/l_H$.
\begin{scalepic}
{H: $k=3$}{II, $\lambda\!\!\sim\!\!1$}{scal03}
\scaleitem{40}{$l_H$, $R_{2,3,4}$}{1}
\scaleitem{180}{$R_1=R_{5,6}=R$}{$l_{II}$, $T^2(R_{5,6})$}
\scaleitem{250}{$R^{3/2}$}{$K3$}
\end{scalepic}

For $k=4$, the type II dual theory provides a perturbative description
alternative to the type I$^\prime$ with $n=2$ extra large transverse
dimensions. For $k=5$, there is no perturbative type II description,
while for $k=6$, the heterotic $E_8\times E_8$ theory is described by a
weakly coupled type IIA with all scales of order $R$ apart one $K3$
direction ($R_I$) which is extra large. This is equivalent to type
I$^\prime$ with $n=1$ extra large transverse dimension. Note that this
case was not found from heterotic $SO(32)$ -- type I duality since the
heterotic $SO(32)$ string is equivalent to $E_8\times E_8$ only up to
T-duality, which cannot be performed when $k=6$ and there are no leftover
dimensions of heterotic size.

\section{ Theoretical implications }

We will now focus on some theoretical implications of the low scale
string scenario. Unless explicitly stated otherwise, we will restrict
ourselves to the context of type I strings.

\subsection {U.V./ I.R. correspondence}

In addition to the open strings describing the gauge degrees of
freedom, consistency of string theory requires the presence of closed strings
associated with gravitons and different kind of moduli fields $m_a$. 

There are two types of extended objects: $D$-branes and orientifolds.
The former are hypersurfaces on which open strings end while the latter are 
hypersurfaces located at fixed points  when acting simultaneously with a
$Z_2$ parity on the transverse  space and world-sheet coordinates.

Closed strings can be emitted by $D$-branes and orientifolds, the lowest
order diagrams being described by a cylindric topology. In this way D-branes
and orientifolds appear as to lowest order classical point-like sources in the
transverse space.  For weak  type-I string coupling this can be 
described by a lagrangian of the form
 
\be
\int d^n x_\perp \; \Bigl[ {1\over g_s^2}
(\partial_{x_\perp}  m_a)^2 +  {1\over g_s} \sum_s f_s(m_a) 
\delta(x_\perp - {x_\perp}_s)\Bigr] \, ,
\label{b}
\ee
where ${x_\perp}_s$ is the location of the source $s$ ($D$-branes and
orientifolds) while $f_s(m_a)$ encodes the coupling of this source to
the moduli $m_a$. As a result while  $m_a$  have constant values in the 
four-dimensional space, their  expectation values will generically vary as a 
function of the transverse coordinates $x_\perp$ of the $n$ directions with 
 size $\sim R_\perp$ large compared to the string length $l_s$. 

Solving the classical equation of motion  for $m_a$ in (\ref{b}) leads to
contributions to the parameters (couplings) on the brane of the low energy
effective action  given by a sum of Green's functions of the form~\cite{ab}:
\be
  {1\over V_\perp}\  \sum_{\vert {p}_\perp\vert < M_{s} }\
{1\over p_\perp^2}\  F({\vec p_\perp})\, ,
\label{tadpole}
\ee 
where $V_\perp={R_\perp}^{d_\perp}$ is the volume of the
transverse space, ${\vec p}_\perp =(m_1/{R_\perp}$$\cdots$
${m_{d_\perp}/{R_\perp}})$ is the transverse momentum exchanged  by
the massless closed string, $F({\vec p_\perp})$ are the
Fourier-transformed to momentum space of derivatives of $f_s(m_a)$. 
An explicit
expression can be given in the simple case of toroidal
compactification with vanishing antisymmetric tensor, where the global 
tadpole cancellation fixes  the number of D-branes to be 32:
\be
F({\vec p_\perp})\sim\left( 32\prod_{i=1}^{d_\perp}{1+(-)^{m_i}\over 2}
-2\sum_{a=1}^{16}{\rm cos}({\vec p_\perp}{\vec x_a})\right)\, ,
\label{tadpole2}
\ee
where  ${\vec p}_\perp =(m_1/{R_\perp}$$\cdots$
${m_{d_\perp}/{R_\perp}})$, the orientifolds are located at the corners of 
the cell 
$[0, \pi R_\perp]^{d_\perp}$ and are responsible for the first term in 
(\ref{tadpole2}), and $\pm{\vec x_a}$ are the transverse
positions of the 32 D-branes (corresponding to Wilson lines of the
T-dual picture) responsible of the second term.

In a compact space where flux lines can not escape to infinity, the Gauss-law 
implies that the total charge, thus global tadpoles, should vanish $F(0)=0$ 
while local tadpoles may not vanish $F({\vec p_\perp}) \neq 0$ for $\vec p
\neq 0$. In that case, obtained for generic positions of the D-branes, the 
tadpole contribution (\ref{tadpole}) leads to the following behavior in the
large radius limit  for the moduli $m_a$:
\be
 m_a ({x_\perp}_s) \sim \cases{ O(R_\perp M_s)\ &for \ \ \ $d_\perp=1$\cr
O(\ln R_\perp M_s)\ \ \ &for \ \ \ $d_\perp=2$\cr
O(1) \ &for\ \ \ \ $d_\perp>2$\cr}\quad ,
\ee
which is dictated by the large-distance behavior of the two-point Green 
function in the $d_\perp$-dimensional transverse space.

There are some important implications of these results:

\begin{itemize} 

\item The tree-level exchange diagram of a closed string can also be seen
as  one-loop exchange of open strings. While from the former point of
view, a long cylinder represents an infrared limit where one computes
the effect of exchanging light closed strings at long distances, in the
second point of view the same diagram is conformally mapped to an
annulus describing the one-loop running in the ultraviolet limit of
very heavy open strings stretching between the two  boundaries of the
cylinder. Thus, from the brane gauge theory point of view, there are
ultraviolet effects that are not cut-off by the string scale $M_s$ but
instead by the winding mode scale $R_\perp M_s^2$.

\item In the case of one large dimension $d_\perp=1$, the corrections are 
linear in $R_\perp$. Such correction appears for instance for the dilaton field
which sits in front of gauge kinetic terms, that drive the
theory rapidly to a strong coupling singularity and, thus, forbid the
size of the transverse space to become much larger than the string length.
It is possible to avoid such large corrections if the
tadpoles cancel locally. This happens when
D-branes are equally distributed at the two fixed points of the
orientifold. 

\item The case $d_\perp=2$ is particularly attractive because it
allows the effective couplings of the brane theory to depend
logarithmically on the size of the transverse space, or equivalently
on $M_P$, exactly as in the case of softly broken supersymmetry at
$M_s$.  Both higher derivative and higher string loop corrections to
the bulk supergravity lagrangian are expected to be small for slowly
(logarithmically) varying moduli. The {\it classical} equations of motion of 
the effective 2d
supergravity in the transverse space are analogous to the
renormalization group equations used to  resum  large corrections to
the effective field  theory parameters with appropriate boundary
conditions.

\end{itemize}

It turns out that low-scale type II theories with infinitesimal string
coupling share many common properties with type I$^\prime$ when
$d_\perp=2$~\cite{ap}. In fact, the limit of vanishing coupling does not
exist due to subtleties related to the singular character of the
compactification manifold and to the non perturbative origin of gauge
symmetries. In general, there are corrections depending logarithmically
on the string coupling, similarly to the case of type I$^\prime$ strings
with 2 transverse dimensions.

\subsection {Unification}

One of the main success of low-energy supersymmetry is that the three
gauge  couplings of the Standard Model, when extrapolated at high
energies assuming the particle content of its $N=1$ minimal
supersymmetric extension (MSSM), meet at an energy scale $M_{\rm
GUT}\simeq 2\times 10^{16}$ GeV. This running is described at the the
one-loop level by: 
\be {1\over g^2_a(\mu)}={1\over g^2}+{b_a\over
4\pi} \ln{M_{\rm GUT}^2\over\mu^2}\, , \ee 
where $\mu$ is the energy
scale and $a$ denotes the 3 gauge group factors of the Standard Model
$SU(3)\times SU(2)\times U(1)$. Note that even in the absence of any
$GUT$ group, if one requires keeping unification of all gauge couplings
then the string relations we discussed in section 4
suggest that the gauge theories arise from the same kind of branes.

Decreasing the string scale below energies of order $M_{GUT}$ is
expected to cut-off the running of the couplings before they meet and
thus spoils the unification. Is there a way to reconcile the apparent
unification with a low string scale?

One possibility is to use power-law running that may
accelerate unification in an energy region where the theory becomes
higher dimensional~\cite{ddg}. Within the effective field theory, the
summation over the KK modes above the compactification scale and below
some energy scale
$E\gg R^{-1}$ yields:
\be
{1\over g^2_a(E)}={1\over g^2_a(R^{-1})}-{b_a^{SM}\over 2\pi}\ln(ER)-
{b_a^{KK}\over 2\pi}\left\{ 2\left( ER-1\right)-\ln(ER)\right\}\, ,
\label{powerev}
\ee
where we considered one extra (longitudinal)
dimension. The first logarithmic term corresponds to the usual 4d
running controlled by the Standard Model beta-functions $b_a^{SM}$, while
the next term is the contribution of the KK tower dominated by the
power-like dependence $(ER)$ associated to the effective multiplicity
of KK modes and controlled by the corresponding beta-functions
$b_a^{KK}$.

Supersymmetric theories in higher dimensions have at least$N=2$
extended supersymmetry thus   the KK excitations form supermultiplets of 
$N=2$. There are two kinds of such supermultiplets, the vector
multiplets containing spin-1 field, a Dirac fermion and 2 real scalars
in the adjoint representation and  hypermultiplets  containing 
an $N=1$ chiral multiplet and its mirror. As the gauge degrees of freedom
are to be identified with bulk fields, their KK excitations will be part of 
$N=2$ vector multiplets. The higgs and matter fields, quarks and leptons, can 
on the other hand be chosen to be either localized without KK 
excitations or instead identified with bulk states with KK excitations forming 
$N=2$ hypermultiplets representations. Analysis of unification with 
the corresponding coefficients  
$b_a^{KK}$  has been performed in \cite{poweruni}.

There are two remarks to be made on this approach: (i) the result
is very sensitive (power-like) to the initial conditions and thus to
string threshold corrections, in contrast to the usual unification based
on logarithmic evolution, (ii) only the case of one extra-dimension
appears to lead to power-like corrections in type I models.

In fact the one-loop corrected gauge couplings in $N=1$ orientifolds are
given by the following expression~\cite{abd}:
\be
{1\over g^2_a(\mu)}={1\over g^2}+s_a m+
{b_a\over 4\pi}\ln{M_I^2\over\mu^2}-\sum_{i=1}^3{b_{a,i}^{N=2}\over 4\pi}
\left\{\ln T_i +f(U_i)\right\}\, ,
\label{thresholds}
\ee
where the first two terms in the r.h.s. correspond to the tree-level
(disk) contribution and the remaining ones are the one-loop (genus-1)
corrections. Here, we assumed that all gauge group factors correspond to
the same type of D-branes, so that gauge couplings are the same to lowest
order (given by $g$). $m$ denotes a combination of the twisted moduli,
whose VEVs blow-up the orbifold singularities and allow the transition to
smooth (Calabi-Yau) manifolds. However, in all known examples, these VEVs
are fixed to $m=0$ from the vanishing of the D-terms of anomalous
$U(1)$'s.

As expected, the one-loop corrections contain an infrared divergence,
regulated by the low-energy scale $\mu$, that produces the usual 4d
running controlled by the $N=1$ beta-functions $b_a$. The last sum
displays the string threshold corrections that receive contributions only
from $N=2$ sectors, controlled by the corresponding $N=2$
beta-functions $b_{a,i}^{N=2}$. They depend on the geometric moduli
$T_i$ and $U_i$, parameterizing the size and complex structure of the
three internal compactification planes. In the simplest case of a
rectangular torus of radii $R_1$ and $R_2$, $T=R_1R_2/l_s^2$ and
$U=R_1/R_2$. The function $f(U)=\ln\left({\rm Re}U|\eta(iU)|^4\right)$
with $\eta$ the Dedekind-eta function; for large $U$, $f(U)$ grows
linearly with $U$.  Thus, from expression (\ref{thresholds}), it follows
that when $R_1\sim R_2$, there are logarithmic corrections (as explained
 for transverse directions to the brane in the previous subsection)
$\sim\ln(R_1/l_s)$, while when $R_1>R_2$, the corrections grow linearly
as $R_1/R_2$. Note that in both cases, the corrections are proportional
to the $N=2$ $\beta$-functions and there no power law corrections in the case 
of more than one large compact dimensions.

Obviously, unification based on
logarithmic evolution requires the two (transverse) radii to be much
larger than the string length, while power-low unification can happen
either when there is one longitudinal dimension a bit larger than the
string scale  ($R_1/R_2\sim R_\parallel/l_s$ keeping $g_s<1$), or
when one transverse direction is bigger than the rest of the bulk.

The most advantageous possibility is to obtain large logarithmic thresholds
depending on two large dimensions transverse to the brane ($d_\perp=2$).
One hopes that such logarithmic
corrections may restore the ``old" unification picture with a GUT
scale given by the winding scale, which for millimeter-size dimensions
has the correct order of magnitude~\cite{cb,ab,admr}. In this way, the
running due to a large desert in energies is replaced by an effective
running due to a ``large desert" in transverse distances from our
world-brane. However, the logarithmic contributions are model
dependent~\cite{abd} and at present there is no compelling explicit
realization of this idea.

\subsection{Supersymmetry breaking and scales hierarchy}

When decreasing the string scale, the question of hierarchy of scales i.e. of
why the Planck mass is much bigger than the weak scale, is translated
into the question of why there are transverse dimensions much larger
than the string scale, or why the string coupling is very
small. For instance for a string scale in the TeV range, 
 From eq.(\ref{treei}) in type I/I$^\prime$ strings, the required
hierarchy $R_\perp/l_I$ varies from $10^{15}$ to $10^5$, when the number
of extra dimensions in the bulk varies from $n=2$ to $n=6$, respectively,
while in type II strings with no large dimensions, the required value of
the coupling $\lambda_{II}$ is $10^{-14}$.

There are two issues that one needs to address:

\begin{itemize}

\item We have seen in section 6.1 that although the string 
scale is very low, there might be large quantum corrections that arise, 
depending on the size of the large dimensions transverse to the brane.
This is as if the UV
cutoff of the effective field theory on the brane is not the
string scale but the winding scale $R_\perp M_I^2$, dual to the large
transverse dimensions and which can be much larger than the string scale.
In particular such correction could spoil the nullification of gauge hierarchy 
that remain the main theoretical motivation of TeV scale strings.

\item Another important issue is to understand the dynamical question on 
the origin of the hierarchy.
 
\end {itemize}

TeV scale strings offer
a solution to the technical (at least) aspect of gauge hierarchy without
the need of supersymmetry, provided there is no effective propagation of
bulk fields in a single transverse dimension, or else closed string
tadpoles should cancel locally. The case of $d_\perp =2$ leads to 
a logarithmic dependence of the effective potential
 on $R_\perp / l_s$ which allows the possible radiative generation  of the 
hierarchy between $R_\perp$ and $l_s$ as for no-scale models. Moreover, it
is interesting to notice that the ultraviolet behavior of the theory is very 
similar with the one with soft supersymmetry breaking at $M_s \sim TeV$.  It
is then natural to ask the question whether there is any motivation leftover
for supersymmetry or not. This bring  us to the problems of the stability of
the new hierarchy and  of the cosmological constant~\cite{aadd}.

In fact, in a non-supersymmetric string theory, the bulk energy density
behaves generically as $\Lambda_{\rm bulk}\sim M_s^{4+n}$, where $n$ is
the number of transverse dimensions much larger than the string length.
In the type I/I$^\prime$ context, this induces a cosmological constant on
our world-brane which is enhanced by the volume of the transverse space
$V_\perp\sim R_\perp^n$. When expressed in terms of the 4d parameters
using the type I/I$^\prime$ mass-relation (\ref{treei}), it is translated
to a quadratically dependent contribution on the Planck mass:
\be
\Lambda_{\rm brane}\sim M_I^{4+n}R_\perp^n\sim M_I^2 M_P^2\, ,
\label{lambda}
\ee
where we used $s=I$. This contribution is in fact the analogue of the
quadratic divergent term Str${\cal M}^2$ in softly broken supersymmetric
theories, with $M_I$ playing the role of the supersymmetry breaking
scale. 

The brane energy density (\ref{lambda}) is far above the (low) string
scale $M_I$ and in general destabilizes the hierarchy that one tries to
enforce. One way out is to resort to special models with broken
supersymmetry and vanishing or exponentially small cosmological
constant~\cite{ks}. Alternatively, one could conceive a different
scenario, with supersymmetry broken primordially on our world-brane
maximally, i.e. at the string scale which is of order of a few TeV. In
this case the brane cosmological constant would be, by construction,
${\cal O}(M_I^4)$, while the bulk would only be affected by
gravitationally suppressed radiative corrections and thus would be almost
supersymmetric~\cite{aadd,ads}. In particular, one would expect the
gravitino and other soft masses in the bulk to be  extremely small
$O(M_I^2/M_P)$. In this case, the cosmological constant induced in the
bulk would be
\be
\Lambda_{\rm bulk}\sim M_I^4/R_\perp^n\sim M_I^{6+n}/M_P^2\, ,
\label{lambdasmall}
\ee
i.e. of order (10 MeV)$^6$ for $n=2$ and $M_I\simeq 1$ TeV.
The scenario of brane supersymmetry breaking is also required in models
with a string scale at intermediate energies $\sim 10^{11}$ GeV (or
lower), discussed in section 4.1. It can occur for 
instance on a
brane distant from our world and is then mediated to us by gravitational
(or gauge) interactions.

In the absence of gravity, brane supersymmetry breaking can
occur in a non-BPS system of D-branes. 
The simplest examples are based on orientifold projections of type IIB,
in which some of the orientifold 5-planes have opposite charge, requiring
an open string sector living on anti-D5 branes in order to cancel the RR
(Ramond-Ramond) charge. As a result, supersymmetry is broken on the
intersection of D9 and anti-D5 branes that coincides with the world
volume of the latter. The simplest construction of this type is a 
$T^4/Z_2$ orientifold with a flip of the $\Omega$-projection (world-sheet
parity) in the twisted orbifold sector. It turns out that several
orientifold models, where tadpole conditions do not admit naive
supersymmetric solutions, can be defined by introducing
non-supersymmetric open sector containing anti-D-branes. A typical
example of this type is the ordinary $Z_2\times Z_2$ orientifold with
discrete torsion. 

The resulting models are chiral, anomaly-free, with vanishing RR tadpoles
and no tachyons in their spectrum~\cite{ads}. Supersymmetry is broken at
the string scale on a collection of anti-D5 branes while, to lowest
order, the closed string bulk and the other branes are supersymmetric. In
higher orders, supersymmetry breaking is of course mediated to the
remaining sectors, but is suppressed by the size of the transverse space
or by the distance from the brane where supersymmetry breaking primarily
occurred. The models contain in general uncancelled NS (Neveu-Schwarz)
tadpoles reflecting the existence of a tree-level potential for the NS
moduli, which is localized on the (non-supersymmetric) world volume of
the anti-D5 branes.

As a result, this scenario implies the absence of supersymmetry on our
world-brane but its presence in the bulk, a millimeter away! The bulk
supergravity is needed to guarantee the stability of gauge hierarchy
against large gravitational quantum radiative corrections.

Explicit examples and methodes of supersymmetry breaking in type I string
theory are described in the Appendix.

\subsection {Electroweak symmetry breaking in TeV-scale strings}

The existence of non-supersymmetric type I string vacua allows us to address
the question of gauge symmetry breaking. 
From the effective field theory point of
view, one expects quadratic divergences in one-loop 
contribution to the masses of scalar fields. It is then important to 
address the following questions: (i) which scale plays the role of the 
Ultraviolet cut-off (ii) could these one-loop corrections be used to 
to generate radiatively the electroweak symmetry breaking, and 
explain the mild hierarchy between the weak and a  string scale at a few TeVs.

A simple framework to address such 
issues is non-supersymmetric ta\-chyon-free $Z_2$ orientifold 
of type IIB superstring compactified to four dimensions on
$T^4/Z_2\times T^2$~\cite{ads}. Cancellation of Ramond-Ramond charges
requires the presence of 32 D9 and 32 anti-D5 (D$\bar5$)
branes. The bulk (closed strings) 
as well as the D9 branes are
$N=2$ supersymmetric while supersymmetry is broken on the world-volume of the
D$\bar5$'s. It is possible \cite{abqhiggs} to compute the effective 
potential involving the
scalars of the D$\bar5$ branes, namely in this simple example 
the adjoints and bifundamentals
of the $USp(16)\times USp(16)$ gauge group. The
resulting potential has a non-trivial minimum which fixes the 
VEV of the Wilson line or, equivalently, the
distance between the branes in the $T$-dual picture. Although the
obtained VEV is of the order of the string scale, the potential
provides a negative squared-mass term when expanded around the
origin. 

In the limit where the radii of the transverse space are large, 
$R_\perp \to\infty$ and for arbitrary longitudinal radius $R_\parallel$,
the result is:
\be
\label{mu2R}
\mu^2(R_\parallel)=-\varepsilon^2(R_\parallel)\, g^2\, M_s^2
\ee
with
\be
\varepsilon^2(R_\parallel) ={1\over 2\pi^2}\int_0^\infty \frac{dl}{\left(2\,
l\right)^{5/2}} 
{\theta_2^4\over 4\eta^{12}}\left(il+{1\over 2}\right) R_\parallel^3
\sum_n n^2 e^{-2\pi n^2R_\parallel^2l}\ .
\label{epsilon2R}
\ee
For the asymptotic value $R_\parallel\to 0$ (corresponding upon T-duality
to a large transverse dimension of radius $1/R_\parallel$),
$\varepsilon(0)\simeq 0.14$, and the effective cutoff for the mass
term at the origin is $M_s$, as can be seen from Eq.~(\ref{mu2R}). At
large $R_\parallel$, $\mu^2(R_\parallel)$ falls off as $1/R_\parallel^2$, 
which is the effective
cutoff in the limit $R_\parallel\to\infty$, in agreement with field theory
results in the presence of a compactified extra
dimension~\cite{SS}. In fact, in the
limit $R_\parallel\to\infty$ an analytic approximation to $\varepsilon(R)$ 
gives:
\be
\varepsilon(R_\parallel)\simeq \frac{\varepsilon_\infty}{M_s\, R_\parallel}\, ,
\qquad\qquad
\varepsilon_\infty^2=\frac{3\, \zeta(5)}{4\, \pi^4}\simeq 0.008\, .
\label{largeR}
\ee

While the mass term (\ref{mu2R}) was computed for the Wilson line it 
also applies, by gauge invariance, to the charged massless fields
which belong to the same representation. By 
orbifolding the previous example, the Wilson line is projected 
away from the spectrum and we are left with  the 
charged massless fields with
quartic tree-level terms and  one-loop negative
squared masses.
By identifying them with the Higgs field we can achieve radiative
electroweak symmetry breaking, and obtain the mild hierarchy
between the weak and string scales in terms of a loop factor. 
More precisely, in the minimal case where there is only one such Higgs
doublet $h$, the scalar potential would
be:
\be
V=\lambda (h^\dagger h)^2 + \mu^2 (h^\dagger h)\, ,
\label{potencialh}
\ee
where $\lambda$ arises at tree-level and is given by an appropriate truncation
of a supersymmetric theory. This property remains valid in any model where
the higgs field comes from an open string with both ends fixed on the same type
of D-branes (untwisted state).
Within the minimal spectrum of the Standard Model,
$\lambda=(g_2^2+g'^2)/8$, with $g_2$ and $g'$ the $SU(2)$ and $U(1)_Y$ gauge
couplings, as in the MSSM. On the other hand,
$\mu^2$ is generated at one loop and can be estimated by
Eqs.~(\ref{mu2R}) and (\ref{epsilon2R}).

The potential (\ref{potencialh}) has a minimum at $\langle h\rangle
=(0,v/\sqrt{2})$, where $v$ is the VEV of the neutral component of the $h$
doublet, fixed by $v^2=-\mu^2/\lambda$. Using the relation of $v$ with the $Z$
gauge boson mass, $M_Z^2=(g_2^2+g'^2)v^2/4$, and the fact that the quartic
Higgs interaction is provided by the gauge couplings as in supersymmetric
theories, one obtains for the Higgs mass a prediction which is the
MSSM value for $\tan\beta\to\infty$ and $m_A\to\infty$:
\be
\label{masa}
M_h=M_Z\ .
\ee
Furthermore, one can compute $M_h$ in terms of the string scale
$M_s$, as $M_h^2=-2\mu ^2=2\varepsilon^2 g^2M_s^2$, or equivalently
\begin{equation}
M_s=\frac{M_h}{\sqrt{2}\, g\varepsilon}
\label{final}
\end{equation}

The determination of the precise value of the string scale suffers from two
ambiguities. The first is the value of the gauge coupling $g$ at $M_s$,
which depends on the details of the model. A second ambiguity concerns the
numerical coefficient
$\varepsilon$ which is in general model dependent.
Varying $R$
from 0 to 5, that covers the whole range of values for a transverse dimension
$1<1/R_\perp <\infty$, as well as a reasonable range for a longitudinal 
dimension
$1<R_\parallel \simlt 5$, one obtains $M_s\simeq 1-5$ TeV. 
In the $R_\parallel\gg 1$ (large longitudinal dimension) region our
theory is effectively cutoff by $1/R_\parallel$ and the Higgs mass is then
related to it by, 
\begin{equation}
\frac{1}{R_\parallel}=\frac{M_h}{\sqrt{2}\, g\,\varepsilon_\infty}\, .
\label{finalR}
\end{equation}
Using now the value for $\varepsilon_\infty$ in the present model,
Eq.~(\ref{largeR}), we find $1/R_\parallel\simgt 1$ TeV.

The tree level Higgs mass has been shown to receive important
radiative corrections from the top-quark sector. For present experimental
values of the top-quark mass, the Higgs mass in Eqs.~(\ref{masa}) and
(\ref{final}) is raised to values around 120 GeV~\cite{higgs}.
In addition there might be large string threshold corrections
in the case of $d_\perp=2$ large transverse dimensions, due to large
logarithms discussed in section 6.1.

\section{ Scenario for studies of experimental constraints }\label{sec:exp}

In order to pursue further, we need to provide the quantum numbers and 
couplings of the relevant light states. In the scenario we consider:
\begin{itemize}

\item Gravitons~\footnote{ Along with gravitons, string models predict the
presence of other very weakly coupled states as gravitini, dilatons,
moduli, Ramond-Ramond fields, etc.} which describe fluctuations of the
metric  propagate in the whole 10- or 11-dimensional space. 

\item In all generality,  gauge bosons propagate on a
$(3+d_\parallel)$-brane, with  $d_\parallel=0,...,6$. However, as we have seen
in the previous sections, a freedom of  choice for the values of the
string and compactification scales requires that gravity and gauge
degrees of freedom live in spaces with different
dimensionalities. This means that $d_{\parallel max} =5$ or 6 for 10-
or 11-dimensional theories, respectively. The value of $d_\parallel$ represents
the number of dimensions felt by KK excitations of gauge bosons.

To simplify the discussion, we will mainly consider the case $d_\parallel=1$ 
where some
of the gauge fields arise from a 4-brane. Since the  couplings  of the
corresponding gauge groups are reduced by  the size of the large dimension
$R_\parallel M_s$ compared to the others, if  $SU(3)$ has KK modes all three
group factors must have. Otherwise it is difficult to reconcile the
suppression of the strong coupling at the string scale with the observed
reverse situation. As a result, there are 5 distinct cases \cite{AAB} 
that we denote
$(l,l,l)$, $(t,l,l)$, $(t,l,t)$, $(t,t,l)$ and $(t,t,t)$, where the three
positions in the brackets correspond to the 3 gauge group factors of the
standard model $SU(3)_c\times SU(2)_w\times U(1)_Y$ and those with $l$ feel
the extra-dimension, while those with $t$ (transverse) do not.

\item The matter fermions, quarks and leptons, are localized on the 
intersection of a
3-brane with the $(3+d_\parallel)$-brane and have no KK excitations along 
the $d_\parallel$ directions.  
Their coupling to KK modes of gauge bosons are given by:
\be
g_n = {\sqrt{2}}\delta^{-{|\vec{\frac{n}{R}}|^2}/{M_s^2}} g
\label{coupling}
\ee
where $\delta >1$ is a model dependent number ($\delta=4$ in the case of
$Z_2$). This is the main assumption in our analysis and limits
derived in the next subsection depend on it. In a more general study it
could be relaxed by assuming that only part of the fermions are localized.
However, if all states are propagating in the bulk, then the KK excitations
are stable and a discussion of the cosmology will be necessary in order to
explain why they have not been seen as isotopes.

Let's denote generically the localized states as $T$ while the bulk states 
with KK momentum $n/R$ by $U_n$, thus the only trilinear allowed couplings are
$g_nTT U_n$ and $g U_n U_m U_{n+m}$ where $g_n$ is given by 
Eq. (\ref{coupling}). Hence because matter fields are localized, 
their interactions do not preserve the 
momenta in the extra-dimension and single KK excitations can be produced.
This means for example that QCD processes $q{\bar q} \rightarrow G^{(n)}$
with $q$ representing quarks and $G^{(n)}$ massive KK excitations 
of gluons are allowed. In contrast, processes such as $G G \rightarrow
G^{(n)}$
are forbidden as gauge boson interactions conserve the internal momenta.

\end{itemize}

\section{Extra-dimensions along the world brane: KK excitations of gauge
bosons}

The experimental signatures of extra-dimensions are of two 
types \cite{ABQ,ABQ2,AAB}:
\begin{itemize}

\item Observation of resonances due to KK excitations. This needs a collider
energy $\sqrt{s} \simgt 1/R_\parallel$ at LHC. 

\item Virtual exchange of the KK excitations which lead to measurable
deviations in cross-sections compared to the standard model prediction.

\end{itemize}
The necessary data needed to evaluate the size of 
these contributions are: the coupling constants given in (\ref{coupling}), 
the KK masses, and the associated widths.
The latter are given by decay rates into standard model fermions $f$:
\begin{equation}
\Gamma\left(X_n\rightarrow f\bar{f}\right)
=g^2_\alpha \frac{M_{\vec n}}{12\pi}C_f(v_f^2+a_f^2)
\label{Gammaf}
\end{equation}
and, in the case of supersymmetric brane there is an additional
contribution from decays  into the scalar superpartners
\begin{equation}
\label{Gammasf}
\Gamma\left(X_n\rightarrow \widetilde{f}_{(R,L)} 
\widetilde{\bar{f}}_{(R, L)}\right)
=g^2_\alpha\frac{M_{\vec n}}{48\pi}C_f(v_f\pm a_f)^2\, ,
\end{equation}
with $C_f = 1$ (3) for color singlets (triplets) and $v_f, a_f$ stand for
the  standard model vector and axial couplings. These widths 
determine the size of corresponding resonance signals and will be 
important when discussing on-shell production of KK excitations.

In the studies of virtual effects, our strategy for extracting
exclusion bounds will depend on the total number of analyzed events.
If it is small then we will consider out of reach compactification
scales  which do not lead to  prediction of at least 3 new events. In
the  case of large number of events, one estimates the  deviation 
from the background fluctuation ($\sim \sqrt{N_T^{\rm SM}(s)}$) by 
computing the  ratio \cite{ABQ,ABQ2,AAB}
\be
\Delta_{T}=\left|{N_T(s)-N_T^{\rm SM}(s) \over \sqrt{N_T^{\rm SM}(s)}}
\right|
\label{deltaT}
\ee
where $N_T(s)$ is the total number of events while $N_T^{\rm SM}(s)$ is 
the corresponding quantity expected from the standard model. These numbers 
are computed using the formula:
\be
N_T =  \sigma  A \int {\cal L} dt
\label{nt}
\ee 
where $\sigma$ is the relevant cross-section, $\int {\cal L} dt$
is the integrated luminosity while $A$ is a suppression factor taking
into account the corresponding efficiency times acceptance factors.

In the next two subsections we derive limits for the case  $(l,l,l)$
where all the gauge factors feel the large extra-dimension.  We will 
return later to the other possibilities.

\subsection {Production at hadron colliders}

At collider experiments, there are three different channels 
$l^+ l^-$, $l^{\pm} \nu$ and dijets where exchange of KK excitations of 
photon+$Z$, $W^{\pm}$ and gluons can produce observable deviations from the 
standard model expectations.

Let's illustrate in details the first case with exchange of neutral bosons.
KK excitations are produced in  
Drell--Yan processes $pp \rightarrow
l^+l^-X$ at the LHC, or $p{\bar p} \rightarrow l^+l^-X$ at the Tevatron, with
$l=e,\mu,\tau$ which  originate from 
the subprocess $q{\bar q}\rightarrow l^+ l^- X$ of centre--of--mass energy $M$.

The two colliding partons take a fraction
\begin{equation}
x_a={M \over \sqrt s}\ e^{y} \quad{\rm and}\quad
x_b={M \over \sqrt s}\ e^{-y}
\end{equation}
of the momentum of the initial proton ($a$) and (anti)proton ($b$), with a
probability described by the quark or antiquark distribution functions\-
$f^{(a)}_{q,\bar q}(x_{a}, M^2)$ and $f^{(b)}_{q,\bar q}(x_{b}, M^2)$. 
The total cross-section, due to the production is given by:
\begin{equation}
\sigma= \sum_{q={\rm quarks}} \int^{\sqrt s }_0 dM \int^{\ln (\sqrt s
/M)}_{\ln (M/\sqrt s)}dy \ g_q (y, M) S_q (y, M) \ ,
\end{equation}
where
\begin{equation}
g_q (y, M)= {M \over 18\pi} x_a x_b \ [f^{(a)}_q (x_a,M^2)
f^{(b)}_{\bar q} (x_b, M^2) + f^{(a)}_{\bar q} (x_a, M^2) f^{(b)}_q (x_b,
M^2)]\ ,
\end{equation}
and
\begin{equation}
S_q (y, M)= 
\sum_{\alpha ,\beta\gamma, Z, KK}g^2_{\alpha}(M) g^2_{\beta}(M) 
{(v^{\alpha}_e v^{\beta}_e+
a^{\alpha}_e a^{\beta}_e)(v^{\alpha}_l v^{\beta}_l + a^{\alpha}_l
a^{\beta}_l) \over (s -m^2_{\alpha} + i\Gamma{_\alpha}
m_{\alpha})(s-m^2_{\beta} - i\Gamma_{\beta} m_{\beta})} \ .
\end{equation}

\begin{figure}[h]
\begin{center}
\includegraphics[width=14cm]{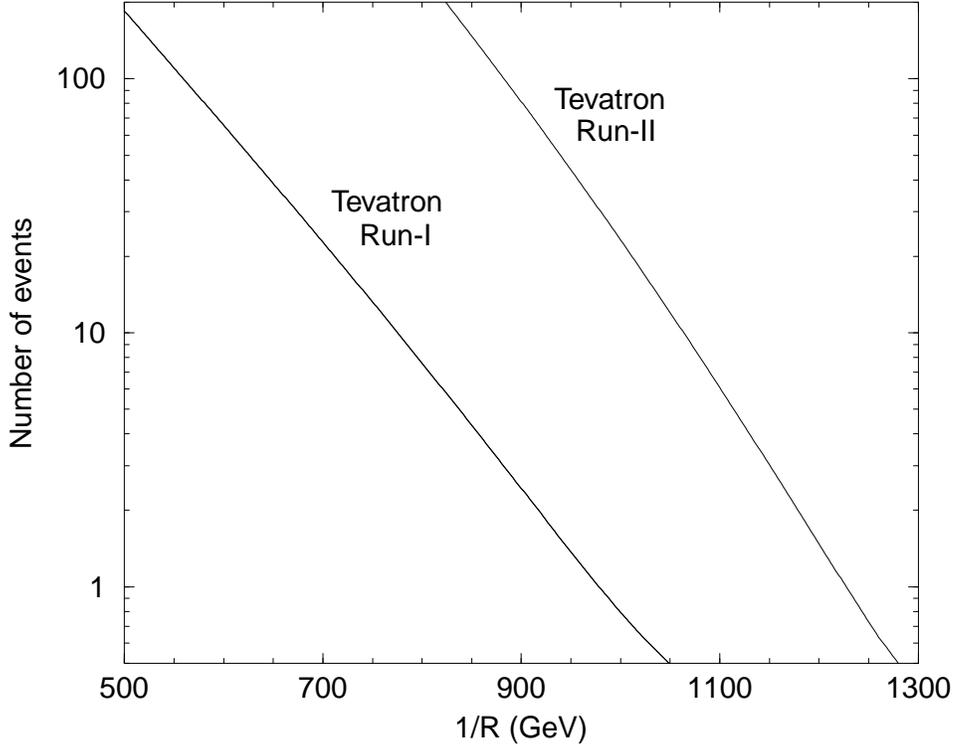}
\end{center}
\caption{\it Number of $l^+ l^-$-pair events with centre--of--mass 
energy above 400 GeV (600 GeV) expected at the Tevatron run-I (run-II) 
with integrated luminosity $\int {\cal L}dt= 110\ pb^{-1}$
($\int {\cal L}dt= 2\ fb^{-1}$) and efficiency times acceptance of
$\sim$ 50\%, as a function of $R^{-1}$.
\label{fig:fig3}}
\end{figure}

 At the Tevatron, the CDF collaboration 
has collected an integrated luminosity $\int {\cal L}dt= 110\ \rm{pb}^{-1}$ 
during the 1992-95 running period. A lower bound on the 
size of compactification scale can be extracted from the absence of candidate 
events at $e^+e^-$ invariant mass above 400 GeV. A similar analysis can be 
carried over for the case of run-II of the Tevatron with a centre--of--mass 
energy $\sqrt{s}=2$ TeV and integrated luminosity 
$\int {\cal L}dt= 2\ fb^{-1}$. The expected number of events at these 
experiments are plotted in  Fig.~\ref{fig:fig3} while the bounds are 
summarized in Table 1 (the factor $A$ in (\ref{nt}) has
 be  taken to be  50 \%) \cite{ABQ2,AAB}.

\begin{table}[t]
\centering
\caption{\it Limits on $R^{-1}_\parallel$ in TeV at present and future
colliders. The luminosity  is given in  fb$^{-1}$.}
\vskip 0.1 in
\begin{tabular}{  | c | c | c | c | l |} 
\hline
  & & & &
\\   Collider & Luminosity &  Gluons  & $W^{\pm}$ & $\gamma + Z$   \\ 
\hline\hline
\mco{5}{|c|}{Discovery of Resonances}   \\ \hline
  LHC     & 100        &  5  & 6  & 6               \\ \hline\hline
\mco{5}{|c|}{Observation of Deviation}   \\ \hline
 LEP 200    &$4\times 0.2$  & - & -  & 1.9 \\ \hline
 TevatronI & $0.11$ &  -  & - & 0.9  \\ \hline 
 TevatronII & 2 &  -  & - & $1.2$ \\ \hline 
  TevatronII  & 20 &  4 & - & 1.3 \\ \hline 
 LHC & 10& 15   & 8.2  & 6.7 \\ \hline
 LHC & 100& 20   & 14 &  12  \\ \hline
  NLC500 & 75& - & - & 8  \\ \hline
  NLC1000 & 200& - & - & 13  \\ \hline
\end{tabular}
\label{tab:1}
\end{table}

\begin{figure}[htb]
\begin{center}
\includegraphics[width=12cm]{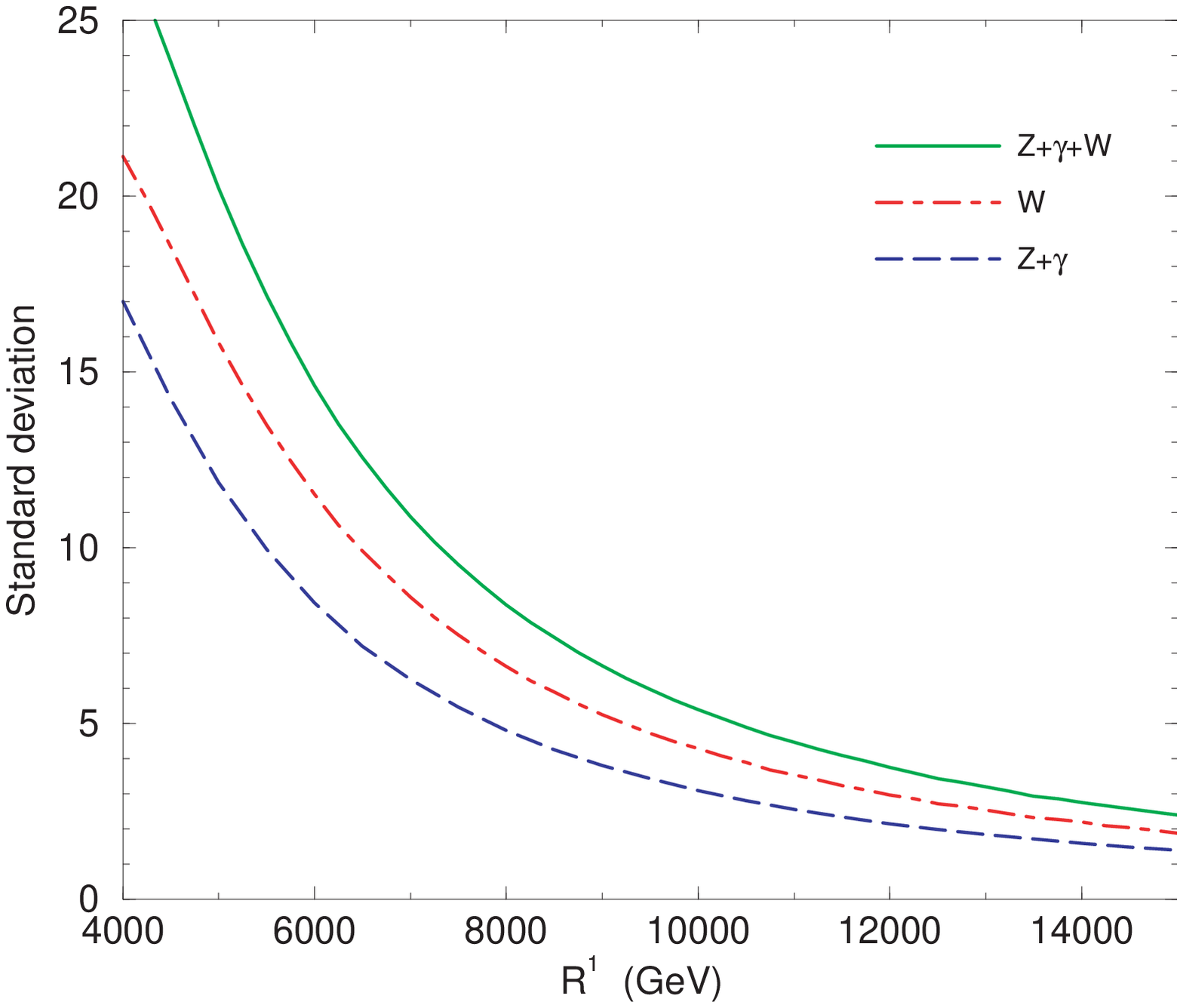}
\end{center}
\caption{\it Number of standard deviation in the number of  $l^+ l^-$
pairs  
and $\nu_l l$ pairs produced from the expected standard model value due to
the presence of one extra-dimension of radius $R$.
\label{fig:fig4}  }
\end{figure}

The most  promising for probing TeV-scale extra-dimensions
are the LHC future experiments at $\sqrt s =14$ TeV with an 
integrated luminosity $\int {\cal L}dt= 100\ fb^{-1}$.  Fig.~\ref{fig:fig4}
shows the expected  deviation from the standard model predictions 
 of the total number of events in the $l^+ l^-$, $l^{\pm} \nu$ due to 
KK excitations $\gamma^{(n)} + Z^{(n)}$ and $W_{\pm}^{(n)}$ respectively.
 The results were obtained by requiring for the dilepton final state 
one lepton to be in  
the central region, $|\eta_l|\le$ 1, the other one having a looser
cut $|\eta_{l^\prime}|\le$ 2.4. Moreover the  lower bound on the transverse 
and invariant mass was chosen to be 400 GeV \cite{AAB,ABQ,ABQ2}.

\begin{figure}[htb]
\begin{center}
\includegraphics[width=12cm]{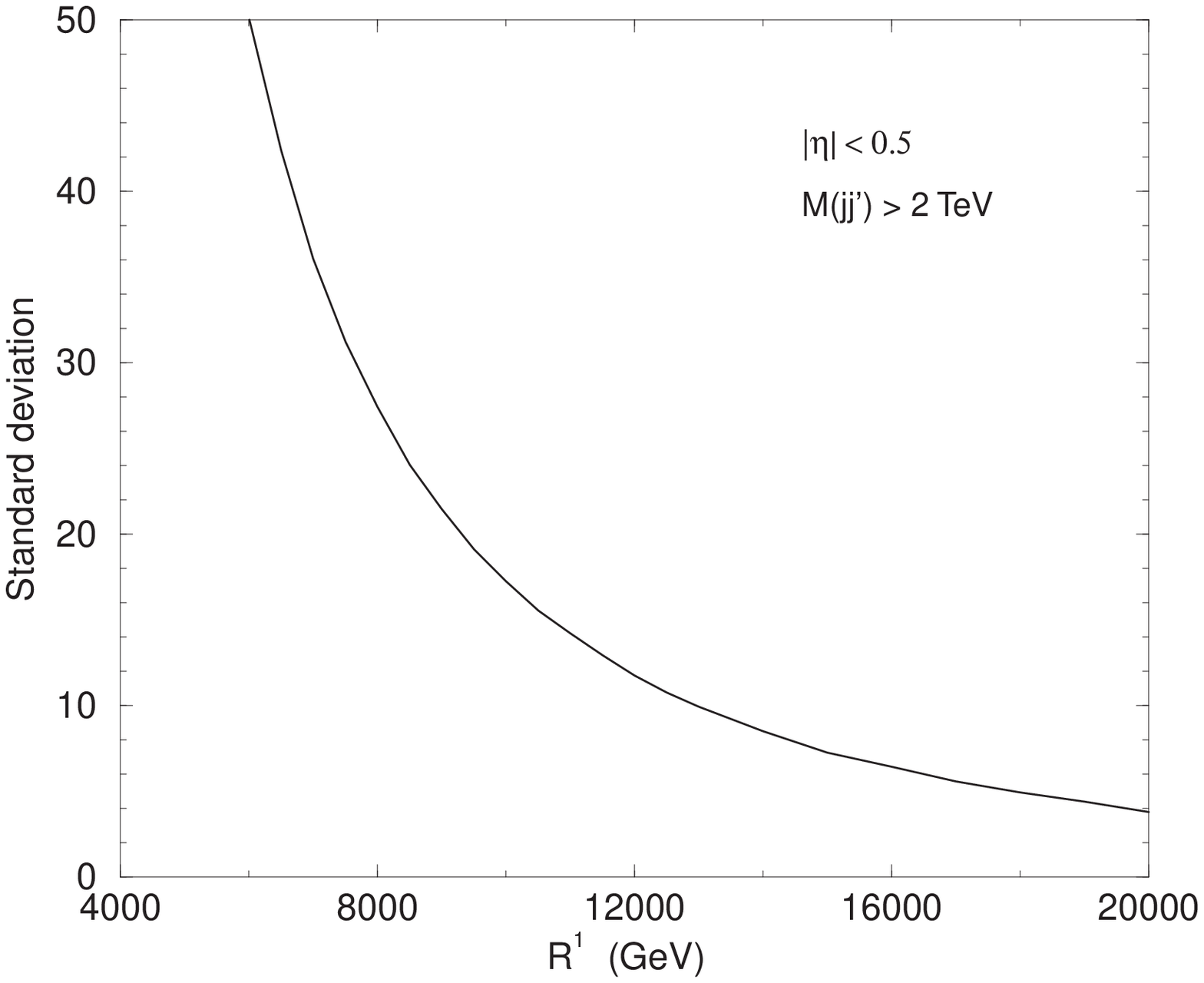}
\end{center}
\caption{\it Number of standard deviation in number of observed dijets
from 
the expected standard model value, due to the presence of a TeV-scale 
extra-dimension of compactification radius $R$.
\label{fig:fig5}}
\end{figure}

In the case of $(l,l,l)$ scenario, looking for an excess of dijet events
due to KK excitations of gluons could be the most efficient 
channel to constrain the size of extra-dimensions. Fig.~\ref{fig:fig5}
shows the corresponding expected   deviation $\Delta_T$ as defined 
in (\ref{deltaT}). This analysis uses 
summation over all jets,
top excluded,  a rapidity cut, $|\eta |\le$
0.5,
on both jets and requirement on the invariant mass to be $M_{jj'}\ge$ 2
TeV, which reduces the SM background and gives the optimal ratio
$S/\sqrt{B}$ especially for large masses \cite{AAB}.

\begin{figure}[h]
\begin{center}
\includegraphics[width=12cm]{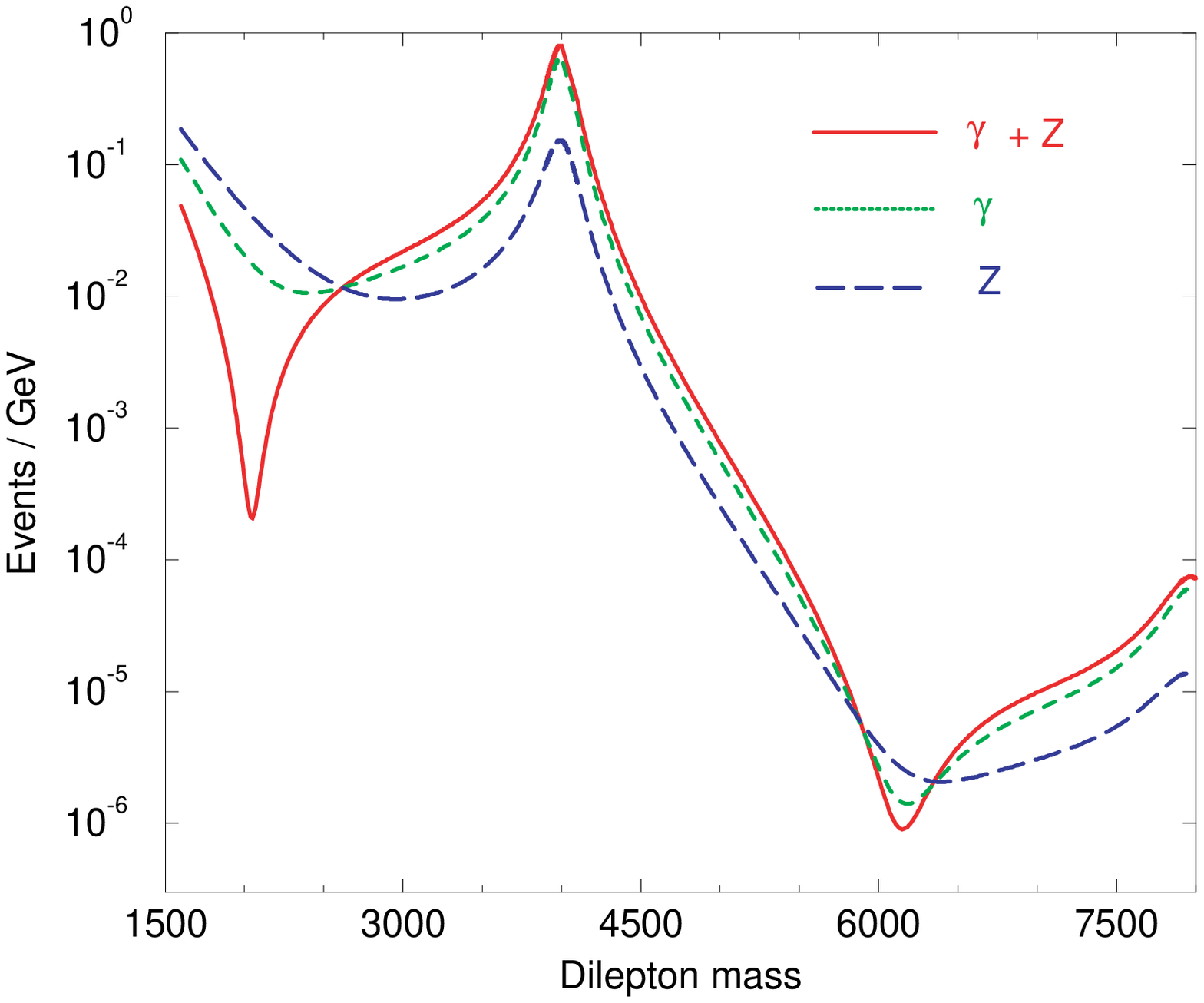}
\end{center}
\caption{\it First resonances in the LHC experiment due to a KK excitation 
of photon and Z  for one extra-dimension at 4 TeV. From highest to lowest: 
excitation of photon+Z, photon and Z boson.
\label{fig:fig6}}
\end{figure}

In addition to these virtual effects, the LHC experiments allow the
production on-shell of KK excitations. The discovery limits for these
KK excitations are given in Table 1. An interesting observation is the
case of excitations $\gamma^{(1)} + Z^{(1)}$ where interferences lead
to a ``deep'' just before the resonance as illustrated in
Fig.~\ref{fig:fig6}

There are some ways to distinguish the corresponding signals from other
possible origin of new physics, such as models with new gauge bosons. In the
case of observation of resonances, one expects three resonances in  the
$(l,l,l)$ case and two in the  $(t,l,l)$ and $(t,l,t)$ cases, located
practically at the same mass value. This property is not shared by most of
other new gauge boson models. Moreover, the heights and widths of the
resonances are directly related to those of standard model gauge bosons in
the corresponding channels.  In the case of virtual effects, these are not
reproduced by a tail of Bright-Wigner shape and a deep is expected just
before the resonance of the photon+$Z$, due to the interference between the
two. However, good statistics will be necessary \cite{ABQ}.

\subsection {High precision data low-energy bounds}

Using the lagrangian describing interactions of the standard 
model states, it is possible to compute all physical observables
in term of few input data. Then one can compare the predictions with 
experimental values.
 
Following ~\cite{scc,Delgado} we will use as  input parameters,
the Fermi constant $G_F=1.166\times
10^{-5}$ GeV$^{-2}$, the fine-structure constant $\alpha=1/137.036$ 
(or $\alpha(M_Z)=1/128.933$) and the
mass of the $Z$ gauge-boson $M_Z=91.1871$ GeV. The observables 
given in Table 2 are then computed with the new lagrangian including 
the contribution of KK excitations. 
%
\begin{table}[h]
\centering
\caption{\it Set of physical observables.  The Standard Model 
predictions are computed for a Higgs mass $M_H=M_Z$ ($M_H=300$ GeV) and a 
top-quark mass $m_t=173\pm 4$ GeV.}
\vskip 0.1 in
\begin{tabular}{||c|c|c||}\hline
Observable & Experimental value & Standard Model prediction \\ \hline
$M_W$ (GeV) & 80.394$\pm$0.042 & 80.377$\pm$0.023 ($-$0.036)\\
$\Gamma_{\ell\ell}$ (MeV) & 83.958$\pm$0.089 & 84.00$\pm$0.03 ($-$0.04)\\
$\Gamma_{had}$ (GeV)& 1.7439$\pm$0.0020 & 1.7433$\pm$0.0016 ($-$0.0005)\\
$A_{FB}^\ell$& 0.01701$\pm$0.00095 & 0.0162$\pm$0.0003 ($-0.0004$)\\
$Q_W$ & $-$72.06$\pm$0.46 & $-$73.12$\pm$0.06 ($+$0.01)\\
$\sum_{i=1}^{3}\left|V_{1i}\right|^2$ & 0.9969$\pm$0.0022 & 1 (unitarity)\\
\hline
\end{tabular}
\label{tab:2}
\end{table}
%
The effects of the latter will be computed 
as a leading order expansion in the small parameter
\begin{equation}
\label{equis}
X=\sum_{n=1}^\infty \frac{2}{n^2}\, \frac{m_Z^2}{M_c^2}=\frac{\pi^2}{3}
{(m_Z R_\parallel)}^2\, ,
\end{equation}
as one expects $ m_Z \ll 1/R_\parallel $.
 
Performing a $\chi^2$ fit, one finds that if the Higgs is a bulk state like the
gauge bosons, $R^{-1} \simgt 3.5$ TeV. Inclusion of $Q_W$ measurement,
which does not give a good agreement with the  standard model itself,
raises the bound to $R^{-1} \simgt 3.9$ TeV~\cite{Delgado}. Different choices 
for localization of matter states and  Higgs  lead to slightly different 
bounds, lying in the 1 to 5 TeV range, and the analysis can  be found  
in ~\cite{Delgado}.

\subsection {One extra dimension for other cases:}

Except for the  $(l,l,l)$ scenario, in all other cases there are no
excitations of gluons and there no important limits from the dijets
channels \cite{AAB}.
The KK excitations  $W_{\pm}^{(n)}$, $\gamma^{(n)}$ and $Z^{(n)}$ are
present 
and lead to the same limits in the $(t,l,l)$ case:  6 TeV for discovery and
15 TeV for the exclusion bounds.
In the $(t,l,t)$ case, only the $SU(2)$ factor feels the
extra-dimension  and the limits are set by KK excitations of $W^{\pm}$
and are again 6 TeV for discovery and 14 TeV for the exclusion bounds.
In the $(t,t,l)$ channel where only $U(1)_Y$ feels 
the extra-dimension the limits are
weaker, the exclusion bound is in fact around 8 TeV.

In addition to these simple possibilities, brane constructions lead often 
to cases where part of $U(1)_Y$
is $t$ and part is $l$, while $SU(3)$ and $SU(2)$ are either $t$ or $l$.
If $SU(3)$ is  $l$ then the bounds come from dijets, if instead $SU(3)$ is  
$t$ and $SU(2)$ is $l$ the limits could come from $W^{\pm}$ while if both are 
$t$ then it will be difficult to distinguish this case 
from a generic extra $U(1)'$.  A good
statistics would be needed to see the deviation in the tail of the
resonance as being due to effects additional to those of a generic  $U(1)'$
resonance.

\subsection {More than one extra dimensions}

The computation of virtual effects of KK excitations involves summing on 
effects of a priori infinite number of tree-level diagrams as terms of the 
form:
\begin{equation}
\sum_{|\vec{n}|} \frac {g^2{(|\vec{n}|)}}{|\vec{n}|^2}
\label{sumdiv}
\end{equation}
arising  from interference  between the exchange of the photon and
$Z$-boson and their KK excitations, with $g^2(|\vec{n}|)$ the KK-mode 
couplings. In the case of one extra-dimension the sum in (\ref{sumdiv})  
converges rapidly and for $R M_s \sim {\cal O}(10)$ the result is not 
sensitive to the value of $M_s$. This allowed us to discuss bounds on only 
one parameter, the scale of compactification. 

In the case of two or more dimensions, Eq. (\ref{sumdiv}) is 
divergent and needs to be regularized using: 
\begin{equation}
g(|\vec{n}|)\sim g\; a_{(|\vec{n}|)}\; 
e^{\frac {-c|\vec{n}|^2}{2R^2 M_s^2}}\, ,
\end{equation}
where $c$ is a constant and  $a_{(|\vec{n}|)}$  takes into account the
normalization of the gauge kinetic terms, as only  the even combination
couples to the boundary. For the case of two  extra-dimensions \cite{ABQ2}
$a_{(0,0)} =1$, $a_{(0,p)} =a_{(q,0)}=\sqrt{2}$ and  $a_{(q,p)} =2$
with $(p,q)$ positive ($>0$) integers. The result  will depend on both
the compactification and string scales.  Other features are that
cross-sections are bigger and  resonances are closer.  The former property
arises because the degeneracy of states  within each mass level
increases with the number of  extra dimensions while the latter
property implies  that more resonances  could be reached by a given hadronic
machine.

\section{Extra-dimensions transverse to the brane world: KK excitations
of gravitons}\label{subsec:miss}

The localization of (infinitely massive) branes in the $(D-d)$ dimensions 
breaks translation invariance  along these directions. Thus, the
corresponding momenta are not conserved: particles, as gravitons, could be
absorbed or emitted from the brane into the $(D-d)$ dimensions. Non
observation of the effects of such processes  allow us to get bounds on the
size of these transverse extra dimensions. In order  to simplify the
analysis, it is usually assumed that among the $D-d$ dimensions $n$ have
very large common radius $R_{\perp } \gg M_s^{-1}$, while the remaining
$D-d-n$ have sizes of the order of the string length.

\subsection{Signals from missing energy experiments}

During a collision of center of mass energy $\sqrt{s}$, there are 
$(\sqrt{s}R_{\perp})^n$ KK excitations of gravitons with mass
$m_{KK\perp}<\sqrt{s}< M_s$, which can be emitted. Each of these states 
looks from the four-dimensional point of view as a massive, quasi-stable, 
extremely weakly coupled ($s/M^2_P$ suppressed) particle that escapes
from the detector. The total effect is a missing-energy cross-section
roughly of order: 
\be
\frac {(\sqrt{s}R_{\perp })^n} {M^2_P} \sim \frac{1}{s} 
{(\frac{\sqrt{s}}{M_s})^{n+2}}
\label{miss1}
\ee

For illustration, the simplest process is the gluon annihilation into a
graviton which escapes into the extra dimensions. The corresponding
cross-section is given by (in the weak coupling limit)~\cite{aadd}:
\be
\sigma(E)\sim {E^n\over M_s^{n+2}}{\Gamma\left(1-2E^2/M_s^2\right)^2
\over\Gamma\left(1-E^2/M_s^2\right)^4}\, ,
\label{sigma}
\ee
where $E$ is the center of mass energy and $n$ the number of extra large
transverse dimensions. The above expression exhibits 3 kinematic regimes
with different behavior. At high energies $E\gg M_s$, it falls off
exponentially due to the UV softness of strings. At energies of the order
of the string scale, it exhibits a sequence of poles at the position of
Regge resonances. Finally, at low energies $E\ll M_s$, it falls off as a
power $\sigma(E)\sim E^n/M_s^{n+2}$, dictated by the effective higher
dimensional gravity which requires the presence of the
$(4+n)$-dimensional Newton's constant $G_N^{(4+n)}\sim l_s^{n+2}$
from eq.(\ref{GN}).

\begin{table}[t]
\centering
\caption{\it Limits on $R_\perp$ in mm from missing-energy
processes.}
\vskip 0.1 in
\begin{tabular}{  | c | c | c | l |} 
\hline
  & & &
\\   Experiment & $R_\perp (n=2)$ & $R_\perp (n=4)$ & $R_\perp (n=6)$ \\ 
\hline\hline
\mco{4}{|c|}{Collider bounds}   \\ \hline

 LEP 2   & $4.8\times 10^{-1}$ & $1.9\times 10^{-8}$  & 
                              $6.8 \times 10^{-11}$ \\ \hline
  Tevatron  &   $5.5 \times 10^{-1}$  & $1.4 \times 10^{-8}$ 
              & $4.1 \times 10^{-11}$ \\ \hline 
  LHC &  $4.5 \times 10^{-3}$   & $5.6\times 10^{-10}$  & 
                              $2.7 \times 10^{-12}$  \\ \hline
  NLC & $1.2\times 10^{-2}$  & $1.2\times 10^{-9}$  & 
                              $6.5 \times 10^{-12}$  \\ \hline\hline
\mco{4}{|c|}{Present non-collider bounds}   \\ \hline
  
SN1987A   &  $3 \times 10^{-4}$   & 
           $1 \times 10^{-8}$ 
                 & $6 \times 10^{-10} $ \\ \hline
COMPTEL &  $5 \times 10^{-5}$   & - & 
                              - \\ \hline
\end{tabular}
\label{tab:exp3}
\end{table}

Explicit computation of these effects leads to the bounds given in 
Table 3, while Fig.~\ref{fig:fig7} shows the
cross-section for graviton emission in the bulk, corresponding to the
process $pp\to jet + gravition$ at LHC, together with the Standard Model
background~\cite{missing}. The results require some remarks:

\begin{figure}[h]
\begin{center}
\includegraphics[width=12cm]{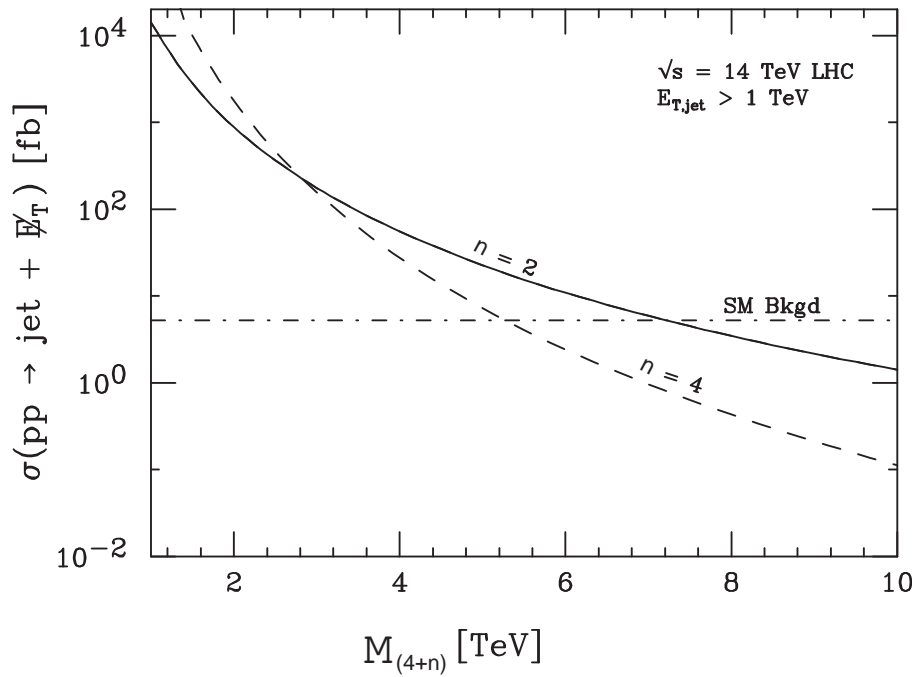}
\end{center}
\caption{\it Missing energy due to graviton emission in the LHC
experiment, as a function of the quantum gravity scale $M_{(4+n)}$ for
$n$ large transverse dimensions, together with the Standard model
baclground.
\label{fig:fig7}}
\end{figure}

\begin{itemize}

\item The amplitude for emission of each of the KK gravitons is taken to be
well approximated by the tree-level coupling of the massless graviton as
derived from General Relativity. Eq.~\ref{coupling} suggests that this is
likely to be a good approximation for $R_{\perp}M_s\gg 1$.

\item The cross-section depends on the size $R_\perp$ of the transverse
dimensions and allows to derive bounds on this {\it physical} scale. 
As it can be seen from Eq.~(\ref{coupling}), transforming these bounds to 
limits on $M_s$ there is an ambiguity on different factors involved, such as
the string  coupling. This is sometimes absorbed in the so called
``fundamental quantum gravity scale $M_{(4+n)}$''.  Generically $M_{(4+n)}$
is bigger than $M_s$, and in some cases, as in type II strings or in
heterotic strings with small instantons, it can be many orders of magnitude
higher than $M_s$. It  corresponds to the scale  where the 
perturbative expansion of string theory seems to break down \cite{veneziano}.

\item There is a particular energy and angular distribution  of the
produced gravitons that arises from the distribution in mass  of KK
states. It might be a smoking gun for the
extra-dimensional nature of such observable signal.

\item For given value of $M_s$, the cross-section for graviton emission
decreases with the number of large transverse dimensions. The effects
are more likely to be observed for the lowest values of $M_s$ and $n$.

\item Finally, while the obtained bounds for $R_\perp^{-1}$ are 
smaller than those that could be checked in table-top experiments probing 
macroscopic gravity at small distances (see next subsection), 
one should keep in mind that 
larger radii are allowed if one relaxes the assumption of isotropy, 
by taking for instance two large dimensions with different radii.

\end{itemize}

In Table 3, we have also included astrophysical and cosmological
bounds. Astrophysical bounds~\cite{add2} arise from the
requirement that the radiation of gravitons should not carry on too much
of the gravitational binding energy released during core collapse of
supernovae. In fact, the measurements of Kamiokande and IMB for SN1987A 
suggest that the main channel is neutrino fluxes.

The best cosmological bound~\cite{COMPTEL} is obtained from requiring that
decay of bulk gravitons to photons do not generate a spike in the energy
spectrum of the photon background measured by the COMPTEL instrument. Bulk 
gravitons  are  expected  to be produced just before
nucleosynthesis due to thermal radiation from the brane. The limits assume
that the temperature was at most 1 MeV as nucleosynthesis begins, and become
stronger if the temperature is increased.

\subsection{Gravity modification and sub-millimeter forces}

Besides the spectacular experimental predictions in particle
accelerators, string theories with large volume compactifications and/or
low string scale predict also possible modifications of gravitation in
the sub-millimeter ran\-ge, which can be tested in ``tabletop" experiments
that measure gravity at short distances. There are two categories of such
predictions:\hfil\\  
(i) Deviations from the Newton's law $1/r^2$ behavior to $1/r^{2+n}$, for
$n$ extra large transverse dimensions, which can be observable for $n=2$ 
dimensions of sub-millimeter size. This case is particularly attractive
on theoretical grounds because of the logarithmic sensitivity of Standard
Model couplings on the size of transverse space, but also for
phenomenological reasons since the effects in particle colliders are
maximally enhanced~\cite{missing}. Notice also the coincidence of this
scale with the possible value of the cosmological constant in the
universe that recent observations seem to support.\hfil\\
(ii) New scalar forces in the sub-millimeter range, motivated by the
problem of supersymmetry breaking discussed in section 6.3, and
mediated by light scalar fields $\varphi$ with
masses~\cite{fkz,iadd,aadd,ads}:
\be
m_{\varphi}\simeq{m_{susy}^2\over M_P}\simeq 
10^{-4}-10^{-2}\ {\rm eV} \, ,
\label{msusy}
\ee
for a supersymmetry breaking scale $m_{susy}\simeq 1-10$ TeV. These
correspond to Compton wavelengths in the range of 1 mm to 10 $\mu$m.
$m_{susy}$ can be either the KK scale $1/R$ if supersymmetry is broken by
compactification~\cite{iadd}, or the string scale if it is broken
``maximally" on our world-brane~\cite{aadd,ads}. A model independent
scalar force is mediated by the radius modulus (in Planck units)
\be
\varphi\equiv\ln R\, ,
\label{varphi}
\ee
with $R$ the radius of the longitudinal or transverse dimension(s),
respectively. In the former case, the result (\ref{msusy}) follows from
the behavior of the vacuum energy density $\Lambda \sim 1/R^4$ for large
$R$ (up to logarithmic corrections). In the latter case, supersymmetry is
broken primarily on the brane only, and thus its transmission to the bulk
is gravitationally suppressed, leading to masses (\ref{msusy}).

The coupling of these light scalars to nuclei can be computed since it
arises dominantly through the radius dependence of $\Lambda_{\rm QCD}$,
or equivalently of the QCD gauge coupling. More precisely, the coupling
$\alpha_\phi$ of the radius modulus (\ref{varphi}) relative to gravity
is~\cite{iadd}:
\be
\alpha_\varphi = {1\over m_N}{\partial m_N\over\partial\varphi}=
{\partial\ln\Lambda_{\rm QCD}\over\partial\ln R}= 
-{2\pi\over b_{\rm QCD}}{\partial\over\partial\ln R}\alpha_{\rm QCD}\, ,
\label{dcoupling}
\ee
with $m_N$ the nucleon mass and $b_{\rm QCD}$ the one-loop QCD
beta-function coefficient. In the case where supersymmetry is broken 
primordially on our
world-brane at the string scale while it is almost unbroken the bulk, the
force (\ref{coupling}) is again comparable to gravity in theories with
logarithmic sensitivity on the size of transverse space, i.e. when there
is effective propagation of gravity in $d_\perp=2$ transverse dimensions.
The resulting forces can therefore be within reach of upcoming
experiments~\cite{price}.

In principle there can be other light moduli which couple with even
larger strengths. For example the dilaton $\phi$, whose VEV determines
the (logarithm of the) string coupling constant, if it does not acquire
large mass from some dynamical supersymmetric mechanism, can lead to a
force of strength 2000 times bigger than gravity~\cite{tvdil}.

\begin{figure}[htb]
\begin{center}
\includegraphics[height=12cm]{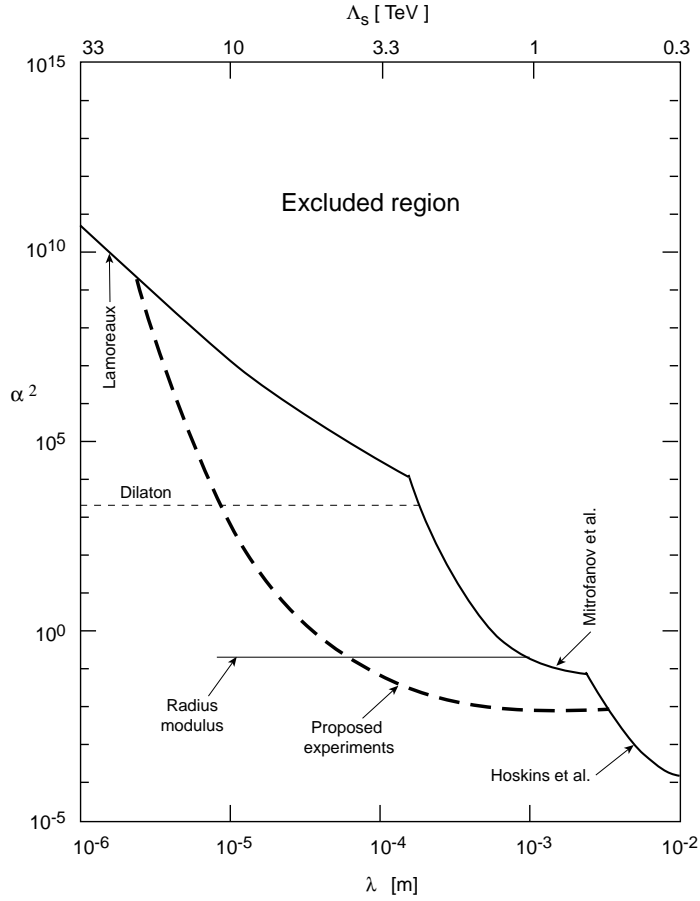}
\end{center}
\caption{\it Strength of the modulus force relative to gravity ($\alpha^2$)
versus  its Compton wavelength ($\lambda$).
\label{fig:fig8}}
\end{figure}

In fig.~\ref{fig:fig8} we depict the actual information from previous,
present and upcoming experiments~\cite{price}. The vertical axis is the
strength, $\alpha^2$, of the force relative to gravity; the horizontal
axis is the Compton wavelength of the exchanged particle; the upper scale
shows the corresponding value of the supersymmetry breaking scale (large
radius or string scale) in TeV. The solid lines indicate the present
limits from the experiments indicated. The excluded regions lie above
these solid lines. Measuring gravitational strength forces at such short  
distances is quite challenging. The most important background is the Van
der Walls force which becomes equal to the gravitational force between
two atoms when they are about 100 microns apart. Since the Van der Walls
force falls off as the 7th power of the distance, it rapidly becomes
negligible compared to gravity at distances exceeding 100 $\mu$m. The
dashed thick line gives the expected sensitivity of the present and
upcoming experiments, which will improve the actual limits by roughly two
orders of magnitude and --at the very least-- they will, for the first
time, measure gravity to a precision of 1\% at distances of $\sim$ 100
$\mu$m.

\section{Dimension-eight operators and limits on the string scale}

At low energies, the interaction of light (string) states is
described by an effective field theory. Non-renormalizable dimension-six
operators are due to the exchange of KK excitations of gauge bosons between
localized states. If these are absent, then there are deviations to the
standard model expectations from dimension-eight operators. There are two
generic sources for such operators: exchange of virtual KK excitations of
bulk fields (gravitons,...) and form factors due to the extended nature of
strings.

The exchange of virtual KK excitations of bulk gravitons is described
in the effective  field theory by an amplitude involving the sum
$\frac {1}{M_P^2}\sum_n \frac {1}{s-\frac{{\vec n}^2}{R_\perp^2}}$. For
$n > 1$, this sum diverges and one cannot compute it in field theory but 
only in a fundamental (string) theory. In analogy with  the case of
exchange of gauge bosons, one expects the string scale to act as a cut-off
with the result: 
\be  
A g_s^2 \frac {T_{\mu \nu}T^{\mu \nu} - \frac {1}{1+d_\perp} 
T_\mu^\mu  T_\nu^\nu} {M_s^4}\, .
\label{QFT}
\ee 
The approximation $A = \log{ \frac{M_s^2}{s}}$ for $d_\perp =2$ and 
$A = \frac{2}{d_\perp-2}$ for $d_\perp > 2$ is usually used for 
quantitative discussions. There are some reasons which might invalidate 
this approximation in general. In fact, the result is 
very much model dependent: in type I string models it reflects the 
ultraviolet behavior of open string one-loop diagrams which are model 
(compactification) dependent.

In order to understand better this issue, it is important to remind that in
type I string models, gravitons and other bulk particles correspond to
excitations of closed strings. Their tree-level exchange is described  by a
 cylinder joining the initial $|B in>$ and final $|B out>$ closed
 strings lying on the brane. This cylinder can be be seen on the other
hand as an open string with one of its end-points describing the
 closed (loop)  string $|B in>$, while the other end draws $|B out>$. In
other words, the cylinder can be seen as an annulus which is a one-loop
 diagram of open strings with boundaries $|B in>$ and $|B out>$. Note that
usually the theory requires the presence of other
weakly interacting closed strings besides gravitons.

More important is that when the gauge degrees of freedom
arise from Dirichelet branes, it is expected that the dominant source
of dimension-eight operators is not the exchange of KK states but
instead the effects of massive open string
oscillators~\cite{AAB,string,abl}. These give rise to
contributions to tree-level scatterings that behave as $g_s s/M_s^4$.
Thus, they are enhanced by a string-loop factor $g_s^{-1}$ compared to
the field theory estimate based on KK graviton exchanges. Although the
precise value of $g_s$ requires a detail analysis of threshold
corrections, a rough estimate can be obtained by taking 
$g_s\simeq\alpha\sim 1/25$, implying an enhancement by one order of
magnitude, and in any case a loop-factor as consequence of perturbation 
theory. 

What is the simplest thing one could do in practice?.  There are some
processes for which there is only one allowed dimension-eight operator; an
example is $f{\bar f}\rightarrow \gamma\gamma$. The coefficient of this
operator can then be computed in terms of $g_s$ and $M_s$.
As a result, in the only framework where computation of such operators is
possible to carry out, one cannot rely on the effects of exchange of KK
graviton excitations in order to derive bounds on extra-dimensions or the
string scale. Instead, one can use the dimension-eight operator arising
from stringy form-factors.

Under the assumption that the electrons arise as open strings on a $D3$-brane, 
and not as living on the intersections of different kind of branes, an 
estimate at the lowest order approximation of string form factor in 
type I  was used in \cite{string}. For instance for $e^+ e^-\to \gamma\gamma$
one has:
\be
 {d \sigma\over d \cos\theta} =
         {d \sigma\over d \cos\theta}\biggr|_{SM} \cdot
\left[ 1 + {\pi^2\over 12 } {ut\over M_S^4}  +
                 \cdots \right]  
\label{photphot}
\ee
while for Bhabha scattering,
it was suggested that
\be
   {d \sigma\over d \cos\theta} (e^-e^+\to e^-e^+)  =
         {d \sigma\over d \cos\theta}\biggr|_{SM} \cdot
\left| {\Gamma(1-\frac {s}{M_s^2}) \Gamma(1-\frac {t}{M_s^2}) \over
\Gamma(1-\frac {s}{M_s^2} -\frac {t}{M_s^2})} \right|^2 \ ,
\label{bhabha}
\ee
where $s$ and $t$ are the Mandelstam kinematic variables.
Using these estimates, present LEP data lead to limits on the string scale 
$M_s \simgt 1$ TeV. This translates into a stronger bound on the size of 
transverse dimension than those obtained from missing energy experiments in 
the cases $d_\perp > 2$. 

On the other hand, when matter fields live on brane
intersections, the presence of dimension-six operators increase the lower
bound on the string scale to 2-3 TeV, independently on the number of large
extra dimensions \cite{abl}.

\section{D-brane Standard Model}

As we discussed in section 6.2, one of the main questions with such a low
string scale is to understand the observed values of the low energy gauge
couplings. One possibility is to have the three gauge group factors of
the Standard Model arising from different collections of coinciding
branes. This is unattractive since the three gauge couplings correspond
in this case to different arbitrary parameters of the model. A second
possibility is to maintain unification by imposing all the Standard Model
gauge bosons to arise from the same collection of D-branes. The large
difference in the actual values of gauge couplings could then be
explained either by introducing power-law running from a few TeV to the
weak scale \cite{ddg,poweruni}, or by an effective logarithmic evolution
in the transverse space in the special case of two large dimensions
\cite{cb,admr,abd}. However, no satisfactory model built along these lines
has so far been presented.

Here, we will discuss a third possibility \cite{akt}, which is
alternative to  unification but nevertheless maintains the prediction of
the weak angle at low energies. Specifically, we consider the strong and
electroweak interactions to  arise from two different collections of
coinciding branes, leading to two different gauge couplings,
\cite{st}. Assuming that the low energy spectrum of the
(non-supersymmetric) Standard Model can be derived by a type I/I$^\prime$
string vacuum, the normalization of the hypercharge is determined in
terms of the two gauge couplings and leads naturally to the right value
of $\sin^2\theta_W$ for a string scale of the order of a few TeV. The
electroweak gauge symmetry is broken by the vacuum expectation values of
two Higgs doublets, which are both necessary in the present context to
give masses to all quarks and leptons.

Another issue of this class of models with TeV string scale is to
understand proton stability. In the model presented here, this is
achieved by the conservation of the baryon number which turns out to be a
perturbatively exact global symmetry, remnant of an anomalous $U(1)$
gauge symmetry broken by the Green-Schwarz mechanism. Specifically, the
anomaly is canceled by shifting a corresponding axion field that gives
mass to the $U(1)$ gauge boson. Moreover, the two extra $U(1)$ gauge
groups are anomalous and the associated gauge bosons become massive  with 
masses of the order of the string scale. Their couplings to the standard
model fields up to dimension five are fixed by charges and anomalies.

\subsection{Hypercharge embedding and the weak angle}

The gauge group closest to the $SU(3)\times SU(2)\times U(1)$ of the Standard
Model one can hope to derive from type I/I$^\prime$ string theory in the above
context is $U(3)\times U(2)\times U(1)$. The first factor arises from three
coincident D-branes (``color" branes). An open string with one end on them is a
triplet under $SU(3)$ and carries the same $U(1)$ charge for all three components.
Thus, the $U(1)$ factor of $U(3)$ has to be identified with {\it gauged} baryon
number. Similarly, $U(2)$ arises from two coincident ``weak" D-branes and the
corresponding abelian factor is identified with {\it gauged} weak-doublet
number. A priori, one might expect that $U(3)\times U(2)$ would be the minimal
choice. However it turns out that one cannot give masses to both up and down quarks in that case.
Therefore,
at least one additional $U(1)$ factor corresponding to an extra D-brane
(``$U(1)$" brane) is necessary in order to accommodate the Standard 
Model. In principle this $U(1)$ brane can be chosen to be independent of the other
two collections with its own gauge coupling. To improve the predictability of the
model, here we choose to put it on top of either the color or the weak D-branes. In
either case, the model has two independent gauge couplings
$g_3$ and $g_2$ corresponding, respectively, to the gauge groups $U(3)$ and
$U(2)$. The $U(1)$ gauge coupling $g_1$ is equal to either $g_3$ or $g_2$.

Let us denote by $Q_3$, $Q_2$ and $Q_1$ the three $U(1)$ charges of $U(3)\times
U(2)\times U(1)$, in a self explanatory notation. Under $SU(3)\times SU(2)\times
U(1)_3\times U(1)_2\times U(1)_1$, the members of a family of quarks and
leptons have the following quantum numbers:
\bea
&Q &({\bf 3},{\bf 2};1,w,0)_{1/6}\nonumber\\
&u^c &({\bf\bar 3},{\bf 1};-1,0,x)_{-2/3}\nonumber\\
&d^c &({\bf\bar 3},{\bf 1};-1,0,y)_{1/3}\label{charges}\\
&L   &({\bf 1},{\bf 2};0,1,z)_{-1/2}\nonumber\\
&l^c &({\bf 1},{\bf 1};0,0,1)_1\nonumber
\eea
Here, we normalize all $U(N)$ generators according to
${\rm Tr}T^aT^b=\delta^{ab}/2$, and measure the corresponding $U(1)_N$ charges
with respect to the coupling $g_N/\sqrt{2N}$, with $g_N$ the $SU(N)$ coupling
constant. Thus, the fundamental representation of $SU(N)$ has $U(1)_N$ charge
unity. The values of the $U(1)$ charges $x,y,z,w$ will be fixed below so that
they lead to the right hypercharges, shown for completeness as subscripts.

The quark doublet $Q$ corresponds necessarily to a massless excitation of an
open string with its two ends on the two different collections of branes. The
$Q_2$ charge $w$ can be either $+1$ or $-1$ depending on whether $Q$
transforms as a $\bf 2$ or a $\bf\bar 2$ under $U(2)$. The antiquark $u^c$
corresponds to fluctuations of an open string with one end on the color
branes and the other on the $U(1)$ brane for $x=\pm 1$, or on other branes in
the bulk for $x=0$. Ditto for $d^c$. Similarly, the lepton doublet $L$
arises from an open string with one end on the weak branes and the other
on the $U(1)$ brane for $z=\pm 1$, or in the bulk for $z=0$. Finally, $l^c$
corresponds necessarily to an open string with one end on the $U(1)$ brane and
the other in the bulk. We defined its $Q_1=1$.

The weak hypercharge $Y$ is a linear combination of the three
$U(1)$'s:\footnote{A study of hypercharge embedding in gauge groups
obtained from M-branes was considered in Ref. \cite{west}.
In the context of Type I ground states such embeddings were considered in 
\cite{ib}.}
\be
Y=c_1 Q_1+c_2 Q_2+c_3 Q_3\, .
\label{Y}
\ee
$c_1=1$ is fixed by the charges of $l^c$ in eq.~(\ref{charges}), while
for the remaining two coefficients and the unknown charges $x,y,z,w$, we obtain
four possibilities:
\bea
c_2 =-{1\over 2}\, ,\, c_3=-{1\over 3}\, ;&
x=-1\, ,\, y=0\, ,\, z=0\, ,\, w=-1\nonumber\\
c_2 ={1\over 2}\, ,\, c_3=-{1\over 3}\, ;&
x=-1\, ,\, y=0\, ,\, z=-1\, ,\, w=1\nonumber\\
c_2 =-{1\over 2}\, ,\, c_3={2\over 3}\,  ;&\!\!\!\!\!\!\!\!\!
x=0\, ,\, y=1\, ,\, z=0\, ,\, w=1\label{solutions}\\
c_2 ={1\over 2}\, ,\, c_3={2\over 3}\, ;&
x=0\, ,\, y=1\, ,\, z=-1\, ,\, w=-1\nonumber
\eea
Orientifold models realizing the $c_3=-1/3$ embedding in the supersymmetric case
with intermediate string scale $M_s\sim 10^{11}$ GeV have been described in 
\cite{ib}.

To compute the weak angle $\sin^2\theta_W$, we use from eq.~(\ref{Y}) that the
hypercharge coupling $g_Y$ is given by~\footnote{The gauge couplings
$g_{2,3}$ are determined at the tree-level by the string coupling and other
moduli, like radii of longitudinal dimensions. In higher orders, they also
receive string threshold corrections.}:
\be
{1\over g_Y^2}={2\over g_1^2}+{4c_2^2\over g_2^2}+
{6c_3^2\over g_3^2}\, ,
\label{gY}
\ee
with $g_1=g_2$ or $g_1=g_3$ at the string scale. On the other hand, with the
generator normalizations employed above, the weak $SU(2)$ gauge coupling is
$g_2$. Thus,
\be
\sin^2\theta_W\equiv{g_Y^2\over g_2^2+g_Y^2}=
{1\over 1+4c_2^2+2g_2^2/g_1^2+6c_3^2g_2^2/g_3^2}\, ,
\label{sintheta}
\ee
which for $g_1=g_2$ reduces to:
\be
\sin^2\theta_W(M_s)=
{1\over 4+6c_3^2g_2^2(M_s)/g_3^2(M_s)}\, ,
\label{sintheta12}
\ee
while for $g_1=g_3$ it becomes:
\be
\sin^2\theta_W(M_s)=
{1\over 2+2(1+3c_3^2)g_2^2(M_s)/g_3^2(M_s)}\, .
\label{sintheta13}
\ee

We now show that the above predictions agree with the experimental value for
$\sin^2\theta_W$ for a string scale in the region of a few TeV. For this
comparison, we use the evolution of gauge couplings from the weak scale $M_Z$ as
determined by the one-loop beta-functions of the Standard Model with three
families of quarks and leptons and one Higgs doublet,
\be
{1\over \alpha_i(M_s)}={1\over \alpha_i(M_Z)}-
{b_i\over 2\pi}\ln{M_s\over M_Z}\ ; \quad i=3,2,Y
\ee
where $\alpha_i=g_i^2/4\pi$ and $b_3=-7$, $b_2=-19/6$, $b_Y=41/6$. We also use
the measured values of the couplings at the $Z$ pole 
$\alpha_3(M_Z)=0.118\pm 0.003$, $\alpha_2(M_Z)=0.0338$, $\alpha_Y(M_Z)=0.01014$
(with the errors in $\alpha_{2,Y}$ less than 1\%). 

In order to compare the theoretical relations for the two cases
(\ref{sintheta12}) and (\ref{sintheta13}) with the experimental value of
$\sin^2\theta_W=g_Y^2/(g_2^2+g_Y^2)$ at $M_s$, we plot in Fig.~1
the corresponding curves as functions of $M_s$. 

%
\begin{figure}[htb]
\hspace{1.5cm}
\includegraphics[width=15cm]{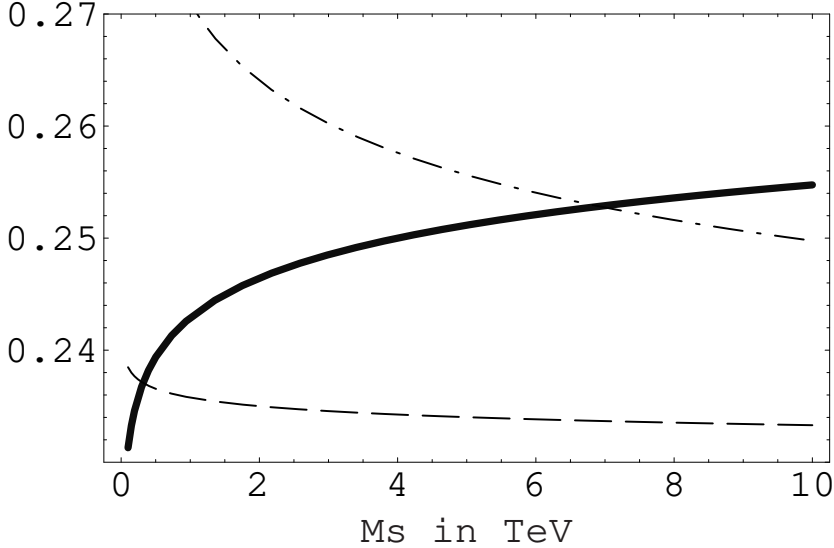}
\vspace{-2.0cm}
\caption{The experimental value of $\sin^2\theta_W$ (thick curve),
together with the theoretical predictions (\ref{sintheta12}) with
$c_3=-1/3$ (dashed line) and (\ref{sintheta13}) with $c_3=2/3$
(dotted-dashed), are plotted as functions of  the string scale $M_s$.}
\label{sin}
\end{figure}

The solid line is the experimental curve. The dashed line is the plot of
the function (\ref{sintheta12}) for $c_3=-1/3$ while the dotted-dashed
line corresponds to the function (\ref{sintheta13}) for $c_3=2/3$. Thus,
the second case, where the $U(1)$ brane is on top of the color branes, is
compatible with low energy data for $M_s\sim 6-8$ TeV and $g_s\simeq 0.9$.
 This selects the last two possibilities of
charge assignments in Eq.~(\ref{solutions}). Indeed, the curve corresponding to
$g_1=g_3$ and $c_3=-1/3$ intersects the experimental curve for $\sin^2\theta_W$
at a scale $M_s$ of the order of a few thousand TeV. Since this value appears to
be too high to protect the hierarchy, it is less interesting and is not shown in
Fig.~1. The other case, where the $U(1)$ brane is on top of the weak branes, is
not interesting either. For $c_3=2/3$, the corresponding curve does not intersect
the experimental one at all and is not shown in the Fig.~1, while the case of
$c_3=-1/3$ leads to $M_s$ of a few hundred GeV and is most likely excluded
experimentally. In the sequel we shall restrict ourselves to the last two
possibilities of Eq.~(\ref{solutions}).

{}From the general solution (\ref{solutions}) and the requirement that the Higgs
doublet has hypercharge $1/2$, one finds the following possible assignments for
it, in the notation of Eq.~(\ref{charges}):
\bea
\label{H}
c_2=-{1\over 2}\ :\qquad &H\ \ ({\bf 1},{\bf 2};0,1,1)_{1/2}\quad
&H'\ \ ({\bf 1},{\bf 2};0,-1,0)_{1/2}\\
c_2={1\over 2}\ :\qquad &{\tilde H}\ \ ({\bf 1},{\bf 2};0,-1,1)_{1/2}\quad
&{\tilde H}'\ \ ({\bf 1},{\bf 2};0,1,0)_{1/2}
\label{Htilde}
\eea
It is straightforward to check that the allowed (trilinear) Yukawa terms
are:
\bea
\label{HY}
c_2=-{1\over 2}\ :\qquad H'Qu^c\ ,\quad H^\dagger Ll^c
\ ,\quad H^\dagger Qd^c\\
c_2={1\over 2}\ :\qquad {\tilde H}' Qu^c\ ,\quad {\tilde H}'^\dagger Ll^c
\ ,\quad {\tilde H}^\dagger Qd^c
\label{HtildeY}
\eea
Thus, two Higgs doublets are in each case necessary and sufficient to give
masses to all quarks and leptons. Let us point out that the presence of
the second Higgs doublet changes very little the curves of Fig.~1 and
consequently our previous conclusions about $M_s$ and $\sin^2\theta_W$.

A few important comments are now in order:

\noindent
(i) The spectrum we assumed in Eq.~(\ref{charges}) does not contain
right-handed neutrinos on the branes. They could in principle arise from
open strings in the bulk. Their interactions with the particles on the
branes would then be suppressed by the large volume of the transverse
space \cite{Rnus}.  More specifically, conservation of the three U(1)
charges allow for the following Yukawa couplings involving the
right-handed neutrino $\nu_R$:
\be
c_2=-{1\over 2}\;:\;\;\; H'~L~{\nu_L}\;\;\;\;\;\;;\;\;\;\;\;\; c_2={1\over 2}\;:\;\;\;
\tilde H~ L ~\nu_R
\ee
These couplings lead to Dirac type neutrino masses between $\nu_L$ from
$L$ and the zero mode of $\nu_R$, which is naturally suppressed by the
volume of the bulk.
 
\noindent
(ii) Implicit in the above was our assumption of three generations
(\ref{charges}) of quarks and lepton in the light spectrum. They can
arise, for example, from an orbifold action along the lines of the model
described in Ref.~\cite{ib}.

\noindent
(iii) From Eq.~(\ref{sintheta13}) and Fig.~1, we find the ratio of the
$SU(2)$ and $SU(3)$ gauge couplings at the string scale to be
$\alpha_2/\alpha_3\sim 0.4$. This ratio can be arranged by an appropriate
choice of the relevant moduli. For instance, one may choose the color and
U(1) branes to be D3 branes while  the weak branes to be D7 branes.
Then, the ratio of couplings above can be explained by choosing the volume
of the four  compact dimensions of the seven branes to be $V_{4}=2.5$ in
string units. This being larger than one is consistent with the picture
above. Moreover it predicts an interesting spectrum of KK states for the
Standard model, different from  the naive choices that have appeared
hitherto: The only Standard Model particles that have KK descendants are
the W bosons as well as the hypercharge gauge boson. However since the
hypercharge is a linear combination of the three U(1)'s the massive U(1)
gauge  bosons couple not to hypercharge but to doublet number.

Another possibility would be to move slightly off the orientifold point
which may be necessary also for other reasons (see discussion next
subsection).

\noindent
(iv) As we have seen that lepton singlet and the u-quark are generated by
strings that must end up in  another brane. This brane must also be
coincident with the rest, in order for the fermions to be light. This
means that these particles will feel the interactions mediated from the
gauge bosons and/or scalars associated with its fluctuations.

\noindent
(v) Finally, it should be stressed that there are some alternative
assignments that may work and these are discussed further in 
\cite{akt}.

\subsection{The fate of $U(1)$'s and proton stability}

The model under discussion has three $U(1)$ gauge interactions
corresponding to the generators $Q_1$, $Q_2$, $Q_3$. From the previous
analysis, the hypercharge was shown to be either one of the two linear
combinations:
\be
Y=Q_1\mp{1\over 2} Q_2+{2\over 3}Q_3\, .
\ee 
It is easy to see that the remaining two $U(1)$ combinations orthogonal
to $Y$ are anomalous. In particular there are mixed anomalies with the
$SU(2)$ and $SU(3)$ gauge groups of the Standard Model.
 
These anomalies are canceled by two axions coming from the closed string
sector, via the standard Green-Schwarz mechanism \cite{gs}.
The mixed anomalies with the non-anomalous hypercharge are also canceled
by dimension five Chern-Simmons  type of interactions \cite{akt}.
The presence of such interactions has so far escaped attention in the
context of string theory.

An important property of the above Green-Schwarz anomaly cancellation
mechanism is that the two $U(1)$ gauge bosons $A$ and $A'$ acquire masses
leaving behind the corresponding global symmetries \cite{gs}. This is in
contrast to what would had happened in the case of an ordinary Higgs
mechanism. These global symmetries remain exact to all orders in type I
string perturbation theory around the orientifold vacuum. This follows
from the topological nature of Chan-Paton charges in all string
amplitudes. On the other hand, one expects non-perturbative violation of
global symmetries and consequently exponentially small in the string
coupling, as long as the vacuum stays at the orientifold point. Once we
move sufficiently far away from it, we expect the violation to become of
order unity. So, as long as we stay at the orientifold point, all three
charges $Q_1$, $Q_2$, $Q_3$ are conserved and since $Q_3$ is the baryon
number, proton stability is guaranteed.

To break the electroweak symmetry, the Higgs doublets in Eq.~(\ref{H}) or
(\ref{Htilde}) should acquire non-zero VEV's. Since the model is
non-supersymmetric, this may be achieved radiatively as we discussed in
subsection 6.4 \cite{abqhiggs}. From Eqs.~(\ref{HY}) and (\ref{HtildeY}),
to generate masses for all quarks and leptons, it is necessary for both
Higgses to get non-zero VEV's. The baryon number conservation remains
intact because both Higgses have vanishing $Q_3$. However, the linear
combination which does not contain $Q_3$, will be broken spontaneously,
as follows from their quantum numbers in Eqs.~(\ref{H}) and
(\ref{Htilde}). This leads to an unwanted massless Goldstone boson of the
Peccei-Quinn type. The way out is to break this global symmetry
explicitly, by moving away from the orientifold point along the direction
of the associated modulus so that baryon number remains conserved.
Instanton effects in that case will generate the appropriate symmetry
breaking couplings  in the potential.

In conclusion, we presented a particular embedding of the Standard Model
in a non-supersymmetric D-brane configuration of type I/I$^\prime$ string
theory. The strong and electroweak couplings are not unified because
strong and weak interactions live on different branes. Nevertheless,
$\sin^2\theta_W$ is naturally predicted to have the right value for a
string scale of the order of a few TeV. The model contains two Higgs
doublets needed to give masses to all quarks and leptons, and preserves
baryon number as a (perturbatively) exact global symmetry. The model
satisfies the main phenomenological requirements for a viable low energy
theory and its explicit derivation from string theory deserves further
study.

\section{Appendix: Supersymmetry breaking in type I strings}

\subsection{Scherk-Schwarz deformations in type-I strings}

Scherk-Schwarz deformations can be introduced in type I strings 
following \cite{kp}, but present a few interesting
novelties, that may be conveniently exhibited referring to
a couple of simple 9D models \cite{ads2}. To this end, we begin by
recalling that, for the type IIB string,
(the fermionic part of) the partition function can be written in the 
compact form
\be
{\cal T} = {|V_8 - S_8 |}^2 \ , \label{ss1}
\ee
resorting to the level-one SO(8) characters
\bea
O_8 = { \vartheta_3^4 + \vartheta_4^4 \over 2 \eta^4} \quad , \quad
V_8 = { \vartheta_3^4 - \vartheta_4^4 \over 2 \eta^4} \quad , \nonumber \\
S_8 = { \vartheta_2^4 - \vartheta_1^4 \over 2 \eta^4} \quad , \quad
C_8 = { \vartheta_2^4 + \vartheta_1^4 \over 2 \eta^4} \quad , \label{ss2}
\eea
where the $\vartheta$ are Jacobi theta functions and $\eta$ is the
Dedekind function. In the usual toroidal reduction, where bosons and
fermions have the momentum modes
\be
p_L = \frac{m}{R} + \frac{n R}{\alpha^\prime} \qquad
p_R = \frac{m}{R} - \frac{n R}{\alpha^\prime} \ , \label{ss3}
\ee
the 9D partition function is
\be
{\cal T} = {|V_8 - S_8 |}^2 \, Z_{mn} \  , \label{ss4}
\ee
where
\be
Z_{mn} \equiv \sum_{m,n} \frac{q^{{\alpha^\prime p_L^2 }/{4}} \
\bar{q}^{{\alpha^\prime p_R^2 }/{4}}}{ \eta \bar{\eta}} \ . \label{ss5}
\ee

A simple modification results in a Scherk-Schwarz breaking of 
space-time supersymmetry. There are actually two inequivalent choices, 
described by
\bea
{\cal T}_1 &=& Z_{m,2n} ( V_8 {\bar V}_8 + S_8 {\bar S}_8 ) +  
Z_{m,2n+1}( O_8
{\bar O}_8 + C_8 {\bar C}_8 )  \nonumber \\
&-& Z_{m+1/2,2n}( V_8 {\bar S}_8 + S_8 {\bar V}_8 ) -
Z_{m+1/2,2n+1}( O_8 {\bar C}_8 + C_8 {\bar O}_8 ) \label{ss6}
\eea
and
\bea
{\cal T}_2 &=& 
Z_{2m,n} ( V_8 {\bar V}_8 + S_8 {\bar S}_8 ) +  
Z_{2m+1,n}(O_8
{\bar O}_8 + C_8 {\bar C}_8 ) \nonumber \\
&-& Z_{2m,n+1/2}( V_8 {\bar S}_8 + S_8 {\bar V}_8 ) -
Z_{2m+1,n+1/2}( O_8 {\bar C}_8 + C_8 {\bar O}_8 ) \ , \label{ss7}
\eea
that may be associated to shifts of the momenta or of the windings of 
the usual ($S_8$) fermionic modes relatively to the usual 
($V_8$) bosonic ones. In both cases modular invariance introduces 
additional sectors, that disappear from the spectrum as the deformation 
is removed, but the two choices are inequivalent, since T-duality 
along the circle interchanges type-IIB and 
type-IIA strings. Both deformed models have tachyon
instabilities at the scale of supersymmetry breaking for the low-lying
modes, $O(1/R)$ for the momentum deformation of eq. (\ref{ss5})
and $O(R/\alpha')$ for the winding deformation of eq. (\ref{ss6}).

The open descendants \cite{cargese} are essentially
determined by the choice of Klein-bottle projection ${\cal K}$ \cite{cc}, while
the other amplitudes ${\cal A}$ and ${\cal M}$ reflect the 
propagation of closed-string modes between boundaries and crosscaps.
In displaying the amplitudes of \cite{ads}, we implicitly confine our 
attention to internal radii such that (closed-string) tachyon instabilities 
are absent, and choose Chan-Paton assignments that remove them 
from the open sectors as well. We also impose some (inessential) 
NS-NS tadpoles, in order to bring the resulting expressions to their 
simplest forms.

Starting from the model of eq. (\ref{ss5}), corresponding to 
{\it momentum shifts}, the additional amplitudes are
\bea 
{\cal K}_1 &=& \frac{1}{2}  \ (V_8 - S_8) \ Z_m \ , \nonumber \\
{\cal A}_1 &=& \frac{n_1^2 + n_2^2}{2} ( V_8 Z_m - S_8 Z_{m + 1/2} )
+ n_1 n_2 ( V_8 Z_{m + 1/2} - S_8 Z_m ) \ ,  \nonumber \\ 
{\cal M}_1 &=& - \frac{ n_1 + n_2 }{2} ( {\hat
V}_8 Z_m - {\hat S}_8 Z_{m + 1/2} ) \ , \label{ss8}
\eea
while the tadpole conditions require that $n_1 + n_2 = 32$. Supersymmetry,
broken in the whole range $R > \sqrt{\alpha'}$,
is recovered asymptotically in the de-compactification limit. This
type-I vacuum, first described in \cite{bd} and interesting in its own right,
describes the type-I string at finite temperature (with Wilson
lines), but includes a 
rather conventional open spectrum, where bosonic and fermionic modes
present the usual $O(1/R)$ Scherk-Schwarz splittings of field-theory models. 

On the other hand, starting from the model of eq. (\ref{ss6}), 
corresponding to {\it winding shifts}, the additional amplitudes are \cite{ads}
\bea  
{\cal K}_2 &=& \frac{1}{2}  \ (V_8 - S_8) \ Z_{2m} + \frac{1}{2} \
(O_8 - C_8) \ Z_{2m+1} \ , \nonumber \\
{\cal A}_2 &=& \biggl( 
\frac{n_1^2 + n_2^2 + n_3^2 + n_4^2}{2} ( V_8 - S_8 )
+ ( n_1 n_3 + n_2 n_4 ) 
( O_8 - C_8 ) \biggr) Z_m \nonumber \\ &+& \biggl( (n_1 n_2 + 
n_3 n_4 ) (V_8  - S_8 )
+ (n_1 n_4 + n_2 n_3 ) (O_8 - C_8 )
\biggr) Z_{m+1/2} \ , \nonumber \\
{\cal M}_2 &=& - \frac{ n_1 + n_2 + n_3 + n_4 }{2} \, {\hat V}_8 \, Z_m +
\frac{ n_1 - n_2 - n_3 + n_4 }{2} \, {\hat S}_8 \, (-1)^m Z_m \ , 
\label{ss9}
\eea
while the tadpole conditions now require that $n_1 + n_2 = n_3 + n_4 = 16$.
Supersymmetry is recovered in the limit of vanishing 
radius $R$, where the whole tower of winding modes present in the 
vacuum-channel amplitudes collapses into additional 
tadpole conditions that eliminate $n_2$ and $n_3$. This is precisely 
the phenomenon of \cite{pw}, spelled out very clearly by these partition 
functions. The resulting open sector, described by
\bea
{\cal A}_2 &=& 
\frac{n_1^2 + n_4^2}{2} ( V_8 - S_8 ) Z_m + n_1 n_4 (O_8 - C_8 ) Z_{m+1/2} 
\ , \nonumber \\
{\cal M}_2 &=& - \frac{ n_1 + n_4 }{2} \, {\hat V}_8 \, Z_m +
\frac{ n_1 + n_4 }{2} \, {\hat S}_8 \, (-1)^m Z_m \ , 
\label{ss10}
\eea
has the suggestive gauge group $SO(16) \times SO(16)$, and
is rather peculiar. In the limit of small breaking $R$, aside from
the ultra-massive $(O,C)$ sector, it contains a conventional $(V,S)$
sector where supersymmetry, {\it exact} for the massless modes, is
effectively broken {\it at the string scale} for the massive ones by the
unpairing of the corresponding Chan-Paton representations. This is
the phenomenon of ``brane supersymmetry'' \cite{ads2}, here present
only for the massless modes. One can then connect, via a sequence of duality
transformations, the $SO(16) \times SO(16)$ gauge group to the two
Horava-Witten walls \cite{hw} of M-theory, with the end result that this
peculiar breaking can be associated to an 11D Scherk-Schwarz deformation. We
are thus facing a simple perturbative description of a phenomenon whose
origin is non-perturbative on the heterotic side.
Several generalizations have been discussed, in six and four dimensions,
both with partial and with total breaking of supersymmetry
\cite{ads2,ks,adds2}.

After suitable T-dualities, these results can be put in 
a very suggestive form: while
the conventional Scherk-Schwarz breaking of ${\cal T}_1$ 
results from shifts {\it parallel} to a brane, the M-theory breaking 
of ${\cal T}_2$ results from shifts {\it orthogonal} to a brane, and is 
ineffective on its massless modes.

\subsection{Brane supersymmetry breaking}
The last phenomenon that we would like to review, 
``brane supersymmetry breaking'' \cite{ads}, solves an old problem in the 
construction of open-string models where, in a number of interesting 
cases, the tadpole conditions have apparently no consistent solution 
\cite{mas}.
The simplest example is provided by the six-dimensional
$T^4/Z2$ reduction where, as in \cite{cc}, the Klein-bottle projection
is reverted for all twisted states. The resulting projected closed
spectrum, described by
\bea
{\cal T} &=& \frac{1}{2} |Q_o + Q_v|^2 \Lambda + \frac{1}{2} |Q_o -
Q_v|^2 {\left|\frac{2 \eta}{\theta_2}\right|}^4 \nonumber \\
&+& \frac{1}{2} |Q_s +
Q_c|^2 {\left|\frac{2 \eta}{\theta_4}\right|}^4
+ \frac{1}{2} |Q_s
Q_c|^2 {\left|\frac{2 \eta}{\theta_3}\right|}^4 \ , \nonumber \\
{\cal K} &=& \frac{1}{4} \left\{ ( Q_o + Q_v ) ( P + W ) -  2 \times
16 ( Q_s + Q_c ) \right\} \ , \label{bsb2}
\eea
contains 17 tensor multiplets and 4 hypermultiplets. Turning on a
 (quantized) NS-NS $B_{ab}$ would lead to similar models with lower 
numbers tensor multiplets, that may be analyzed in a similar 
fashion \cite{carloB}. In writing eq. (\ref{bsb2}), where
$\Lambda$ is the whole Narain lattice sum, while $P$ and $W$ 
are its restrictions to only momenta or only windings, we have
resorted to the supersymmetric combinations of SO(4) characters 
\bea
Q_o &=& V_4 O_4 - C_4 C_4 \quad , \quad Q_v = O_4 V_4 - S_4 S_4 \ ,
\nonumber \\
Q_s &=& O_4 C_4 - S_4 O_4 \quad , \quad Q_c = V_4 S_4 - C_4 V_4  \ .
\label{bsb1}
\eea 
The reversal of the Klein-bottle projection for the twisted states 
changes the relative sign of the crosscap contributions for N and D strings
or, equivalently, the relative charge of the O5 planes relative to the
O9 ones. This is clearly spelled out 
by the terms at the origin of the lattices,
\be
\tilde{\cal K}_0 = \frac{2^5}{4} \biggl\{ Q_o \biggl( \sqrt{v}  \pm
\frac{1}{\sqrt{v}}\biggr)^2 + Q_v \biggl( \sqrt{v}  \mp
\frac{1}{\sqrt{v}}\biggr)^2 \biggr\} \ , \label{bsb3}
\ee
where the upper signs refer to the standard choice, while the lower ones
refer to the reverted Klein bottle of eq. (\ref{bsb2}). In the
latter case one is forced to cancel a {\it negative} background O5 charge,
and this can be achieved introducing antibranes in the vacuum configuration.
The corresponding open sector \cite{ads}
\bea
{\cal A} &=& \frac{1}{4} \biggl\{ (Q_o + Q_v) ( N^2 P  + D^2 W ) + 
2 N D (Q'_s + Q'_c) {\biggl(\frac{\eta}{\theta_4}\biggr)}^2  \label{a10}
\label{a13} \\
&+& (R_N^2 + R_D^2) (Q_o - Q_v) {\biggl(\frac{2
\eta}{\theta_2}\biggr)}^2 + 2 R_N R_D ( - O_4 S_4 - C_4 O_4 + V_4 C_4
+ S_4 V_4 ){\biggl(\frac{
\eta}{\theta_3}\biggr)}^2 \biggr\} \,  \nonumber \\
{\cal M} &=& - \frac{1}{4} \biggl\{ N P ( \hat{O}_4
\hat{V}_4  + \hat{V}_4 \hat{O}_4  - \hat{S}_4 \hat{S}_4
- \hat{C}_4 \hat{C}_4 ) -  D W ( \hat{O}_4
\hat{V}_4  + \hat{V}_4 \hat{O}_4  + \hat{S}_4 \hat{S}_4
+ \hat{C}_4 \hat{C}_4 ) \nonumber \\ &-&\!\!\!\!\! N( 
\hat{O}_4 \hat{V}_4 \!-\! \hat{V}_4 \hat{O}_4 \!-\! \hat{S}_4 \hat{S}_4
\!+\! \hat{C}_4 \hat{C}_4 )\left(
{2{\hat{\eta}}\over{\hat{\theta}}_2}\right)^2  \!\!+\! D( \hat{O}_4
\hat{V}_4 \!-\! \hat{V}_4 \hat{O}_4 \!+\! \hat{S}_4 \hat{S}_4
\!-\! \hat{C}_4 \hat{C}_4)\left(
{2{\hat{\eta}}\over{\hat{\theta}}_2}\right)^2  \biggr\} \ \nonumber
\eea
results from a combination of D9 branes and D5 antibranes.
Supersymmetry is broken on the
D$\bar{5}$ branes, and indeed the amplitudes involve the new 
characters $Q'_s$ and $Q'_c$, corresponding to a chirally flipped 
supercharge, that may
be obtained from eq. (\ref{bsb1}) upon the interchange of $S_4$ and $C_4$,
as well as other non-supersymmetric combinations.
The tadpole conditions determine the gauge group 
$[ SO(16) \times SO(16) ]_9 \times  [ USp(16) \times USp(16) ]_5$, and the
$99$ spectrum is supersymmetric, with (1,0) vector
multiplets for the $SO(16) \times SO(16)$ gauge group and a
corresponding hypermultiplet in the representations ${\bf\!
(16,16,1,1)}$.
On the other hand, the ${\bar 5} {\bar 5}$ DD spectrum is
non supersymmetric, and contains,
aside from the $[ USp(16) \times USp(16) ]$ gauge vectors, quartets
of scalars in ${\bf (1,1,16,16)}$, right-handed Weyl fermions in
${\bf (1,1,120,1)}$, ${\bf (1,1,1,120)}$ and left-handed Weyl fermions in 
${\bf (1,1,16,16)}$.
Finally, the ND sector, also non supersymmetric, comprises doublets of
scalars in ${\bf (16,1,1,16)}$ and in ${\bf (1,16,16,1)}$, and additional
(symplectic) Majorana-Weyl fermions in ${\bf
(16,1,16,1)}$ and ${\bf (1,16,1,16)}$. These fields are a peculiar feature
of six-dimensional space time, where the fundamental Weyl fermion, a 
spinor of $SU^*(4)$, 
is psudoreal, and can thus be subjected to a
Majorana condition if this is supplemented by the
conjugation in a pseudoreal representation.
All irreducible gauge and gravitational
anomalies cancel, while the residual anomaly polynomial 
requires a generalized Green-Schwarz mechanism \cite{gs} 
with couplings more general than those found in supersymmetric models.

It should be appreciated that the resulting
non-BPS configuration of branes and anti-branes have 
no tachyonic excitations, while the branes themselves experience no mutual
forces. Brane configurations of this type have received some attention lately
\cite{sen},
and form the basis of earlier constructions of non-supersymmetric type I
vacua \cite{nonsusy} and of their tachyon-free reductions \cite{carlo}.
As a result, the contributions to the vacuum energy, localized on
the antibranes, come solely from the M\"obius contribution amplitude. 
The resulting
potential, determined by uncancelled NS-NS tadpoles, is
\be
V_{\rm eff}=c{e^{-\phi_6}\over{\sqrt v}}=ce^{-\phi_{10}}
={c\over g_{\rm YM}^2}\, , \label{a18}
\ee
where $\phi_{10}$ is the 10D dilaton, that determines the Yang-Mills coupling
$g_{\rm YM}$ on the D5 branes, and $c$ is some {\it positive} numerical 
constant. This potential (\ref{a18}) is clearly localized on the D5 branes, 
while the D5 brane contribution to the
vacuum energy is positive, consistently with the
interpretation of this mechanism as global supersymmetry breaking.
One would also expect that, in the limit of vanishing 
D5 coupling, supersymmetry
be recovered, at least from the D9 viewpoint. While not true 
in six dimensions,
due to the peculiar chirality flip that we have described, this 
expectation is actually realized after compactification to four dimensions,
with suitable subgroups of the antibrane gauge group realized as internal
symmetries.

Several generalizations of this model have been discussed in \cite{ads}. 
These include the possibility of
allowing the simultaneous presence of branes and antibranes of the
same type, still in tachyon-free combinations, that extend 
the construction of \cite{sug}. This more general setting has the
amusing feature of leading
to the effective stabilization of some geometric moduli, while some of
the resulting models, 
related to the $Z_3$ orientifold of \cite{abpss}, have interesting
three-family spectra, of some potential interest for phenomenology.

\section*{Acknowledgments}
This work was partly supported by the European Commission under
TMR contract  ERBFMRX-CT96-0090, RTN contract HPRN-CT-2000-00148 and 
INTAS contract 99-0590. Variations of these lectures were also given at the
``LNF-INFN Spring School in Nuclear, Subnuclear and Astropartcle Physics",
Frascati, Italy, 15-20 May 2000, at the ``Workshop on Phenomenology of Extra
Dimensions", Glasgow, UK, 31 May-1 June 2000, at the ``NATO ASI school on
Recent Developments in Particle Physics and Cosmology",	Cascais, Portugal,
26 June-7 July 2000, at the ``38th Course on Theory and Experiment Heading
for New Physics", Erice, Italy, 27 August-5 September 2000, and at the ``RTN
Workshop on the Quantum Structure of Spacetime",	Berlin, Germany, 4-10
October 2000.

\end{document}